\documentclass[aps,pre,reprint,superscriptaddress]{revtex4-1}
\usepackage{graphicx}
\usepackage{amsmath}
\usepackage{amssymb}
\usepackage{bm}
\usepackage{physics}
\usepackage{subfigure}
\usepackage[colorlinks=true,linkcolor=blue,urlcolor=blue,citecolor=blue]{hyperref}
\usepackage{times}

\begin{document}
\title{Landau-Zener transitions in a fermionic dissipative environment}
\author{Ruofan Chen}
\affiliation{Science and Technology on Surface Physics and Chemistry
  Laboratory, Mianyang 621908, China}
\date{\today}

\begin{abstract}
  We study Landau-Zener transitions in a fermionic dissipative
  environment where a two-level (up and down states) system is coupled
  to two metallic leads kept with different chemical potentials at
  zero temperature. The dynamics of the system is simulated by an
  iterative numerically exact influence functional path integral
  method. In the pure Landau-Zener problem, two kinds of transition
  (from up to down state and from down to up state) probability are
  symmetric. However, this symmetry is destroyed by coupling the
  system to the bath. In addition, in both kinds of transitions, there
  exists a nonmonotonic dependence of the transition probability on
  the sweep velocity; meanwhile nonmonotonic dependence of the
  transition probability on the system-bath coupling strength is only
  shown in one of them. As in the spin-boson model, these phenomena
  can be explained by a simple phenomenological model.
\end{abstract}
\maketitle

\section{Introduction}
In physics and chemistry, it is ubiquitous that a quantum system can
be effectively described by two-level systems (TLSs). The simplest
example is a particle of total spin \(\frac{1}{2}\) under an external
magnetic field, which can be called an ``intrinsically'' two-level
system. A more common situation is that a system has continuous
degrees of freedom which are associated with a potential with two
minima \cite{leggett1987-dynamics,weiss1993-quantum}. In 1927, Hund
\cite{hund1927-on} first introduced the quantum tunneling effect when
describing the intramolecular rearrangement in ammonia molecules. Soon
after, Oppenheimer \cite{oppenheimer1928-three} used the tunneling
effect to explain the ionization of atoms in strong electric
fields. Since then quantum tunneling in isolated TLSs under external
driving has been widely studied. A well-known example is the so-called
Landau-Zener problem where an isolated TLS undergoes a time-dependent
energy sweep. In such a model, the final transition between states of
the TLS is called the Landau-Zener (LZ) transition, which was first
solved independently by Landau \cite{landau1932-on}, Zener
\cite{zener1932-non}, St\"{u}ckelberg \cite{stueckelberg1932-theorie}
and Majorana \cite{majorana1932-atomi} in 1932.

As one of the most fundamental phenomena in quantum physics, the LZ
transition plays an important role in various fields such as quantum
chemistry \cite{May2011-charge}, atomic and molecular physics
\cite{thiel1990-landau,harmin1994-incoherent,xie2017-accuracy}, solid
state artificial atoms
\cite{sillanpaa2006-continuous,berns2008-amplitude}, spin flips in
nanomagnets \cite{raedt1997-theory,wernsdorfer1999-quantum}, quantum
optics \cite{spreeuw1990-classical,bouwmeester1995-observation},
Bose-Einstein condensates
\cite{witthaut2006-towards,zenesini2009-time,olson2014-tunable},
quantum information and computation
\cite{ankerhold2003-enhancement,izmalkov2004-observation,wubs2005-landau,oliver2005-mach,wei2008-controllable,fuchs2011-quantum},
and Landau-Zener-St\"{u}ckelberg interferometry
\cite{shytov2003-landau,izmalkov2008-consistency,shevchenko2010-landau,neilinger2016-landau,chatterjee2018-silicon,wang2018-landau}.

For isolated TLSs, Landau-Zener transitions can be solved exactly
\cite{landau1932-on,zener1932-non,stueckelberg1932-theorie,majorana1932-atomi,kayanuma1984-nonadiabatic,grifoni1998-driven,wittig2005-landau}.
However, this is no longer the case when taking the environment into
consideration \cite{gefen1987-zener,ao1989-influence,ao1991-quantum}
except for some limiting cases. How the environment affects the
Landau-Zener transition has continuously attracted considerable
attentions over the decades. Kayanuma \cite{kayanuma1984-nonadiabatic}
proposed a simple stochastic model having a diagonal energy
fluctuating term and gave the analytic LZ transition probability in
the rapid fluctuation limit. Gefen \emph{et al.}
\cite{gefen1987-zener} gave a qualitative indication on how the LZ
transition be affected by the environment. Ao and Rammer
\cite{ao1989-influence,ao1991-quantum} studied the LZ transition with
an Ohmic heat bath and they found that at zero temperature in the
limits of very fast and very slow sweeps the transition probability is
the same as in the absence of the bath, which was confirmed by
numerical calculations
\cite{kayanuma1998-nonadiabatic,kobayashi1999-non}. Wubs \emph{et al.}
\cite{wubs2005-landau} investigated the influence of a classical
radiation field on the LZ transition and obtained analytical results
in the limits of large and small frequencies within a rotating wave
approximation. Later they \cite{wubs2006-gauging} gave an exact LZ
transition probability for a qubit with linear coupling to a bosonic
bath at zero temperature and proposed to use the LZ transition to make
qubits as bath detectors. Saito \emph{et al.}
\cite{saito2007-dissipative} studied the LZ transition in a qubit
coupled to bosonic and spin bath respectively at zero temperature and
discussed their bath-specific and universal behaviors. Nalbach and
Thorwart \cite{nalbach2009-landau} studied the LZ transition in a
bosonic dissipative environment by means of an iterative numerically
exact influence functional path integral method, and they discover a
nonmonotonic dependence of the transition probability on the sweep
velocity which can be explained by a simple phenomenological
model. Whitney \emph{et al.} \cite{whitney2011-temperature} found that
the Lamb shift of the environment exponentially enhances the coherent
oscillation amplitude in the LZ transition. Haikka and M\o{}lmer
\cite{haikka2014-dissipative} studied the LZ transition when the
system is subjected to continuous probing of the emitted radiation and
they found the measurement back action on the system leads to
significant excitation. Arceci \emph{et al.}
\cite{arceci2017-dissipative} revisited the issue of thermally
assisted quantum annealing by a detailed study of the dissipative LZ
problem in the presence of a Caldeira-Leggett bath of harmonic
oscillators. Huang and Zhao \cite{huang2018-dynamics} employed the
Dirac-Frenkel time-dependent variation to examine dynamics of the LZ
problem with both diagonal and off-diagonal qubit-bath coupling.

Till now most studies of the effect of environment on the LZ
transition have focused on spin-boson systems where the environment is
described as a bath of harmonic oscillators. The effect of a fermionic
environment is much less well understood. In this article, we employ
an iterative numerically exact influence method
\cite{makarov1994-path,makri1995-numerical,weiss2008-iterative,segal2010-numerically}
to study LZ transitions in a fermionic environment where a TLS is
coupled to two metallic leads kept with different chemical potentials
at zero temperature. Such a method allows us to include nonadiabatic
and non-Markovian effects and is well suited for real-time dynamics
simulation of quantum dots.

In the pure LZ transition problem, two kinds of transition (from up to
down state and from down to up state) probabilities are
symmetric. Whether the spin is initially prepared in up state or down
state, the final probability that it transits to another state is the
same. According to our simulations, this is no longer the case when
the system is coupled to the leads. In addition, a nonmonotonic
dependence of the transition probability on the sweep velocity exists
in both kinds of transition, while nonmonotonic dependence of the
system-bath coupling strength is only shown in one of them. These
phenomena can be explained by a simple phenomenological model as in
the spin-boson model. This nonmonotonic dependence can be understood
as a nontrivial competition between relaxation caused by the
environment and LZ driving, and it may be useful for optimal control
problems.

This article is organized as follows. The details of the model and a
quick survey of the method are given in Sec. \ref{sec:model}. The
simulation results and discussions are shown in
Sec. \ref{sec:result}. Finally, some concluding remarks are given in
Sec. \ref{sec:conclusion}.

\section{Model and Method}
\label{sec:model} 

We consider a spin-fermion system with the time-dependent Hamiltonian
\begin{equation}
  H(t)=H_S(t)+H_B+H_{SB},
\end{equation}
where the system Hamiltonian \(H_S(t)\) is the standard LZ Hamiltonian
for an isolated TLS for which
\begin{equation}
  H_S(t)=\frac{vt}{2}\sigma_z+\frac{\Delta_0}{2}\sigma_x
  \label{eq:lz-hamiltonian}
\end{equation}
with the tunneling amplitude \(\Delta_0\) and the sweep velocity
\(v\). Throughout this article we set \(\hbar=k_B=1\) and use
dimensionless quantities. The value of \(\Delta_0\) is set to
\(\Delta_0^2=0.1\) and the value of \(v\) is kept
positive. \(\sigma_x\) and \(\sigma_z\) are Pauli matrices, and
diabatic states are the eigenstates of \(\sigma_z\) (up
$\ket{\uparrow}$ and down $\ket{\downarrow}$ states). When
\(t\to\pm\infty\), the diabatic states coincide with the momentary
eigenstates of the LZ Hamiltonian.

The bath Hamiltonian \(H_{B}\) describes two independent free fermionic
leads (\(\alpha=L,R\) for left and right lead) for which
\begin{equation}
  H_B=\sum_{\alpha k}\varepsilon_k c^{\dag}_{\alpha k}c_{\alpha k},
\end{equation}
where the operator \(c^{\dag}_{\alpha k}\) (\(c_{\alpha k}\)) creates
(annihilates) an electron in the \(\alpha\)th lead with state \(k\).
These two leads are kept with a chemical potential difference
\(\Delta\mu\) at zero temperature. Here we suppose the chemical
potential of the left lead \(\mu_L\) is higher than that of the right
lead \(\mu_R\), and the bandwidth of both leads is taken to be
\(D=3\). The middle of the band is set as the zero energy point, and
\(\mu_L,\mu_R\) are symmetrically placed on two sides of it, i.e.,
\(\mu_L=\frac{1}{2}(D+\Delta\mu)\) and
\(\mu_R=\frac{1}{2}(D-\Delta\mu)\).

\begin{figure}[htbp]
  \centerline{\includegraphics[]{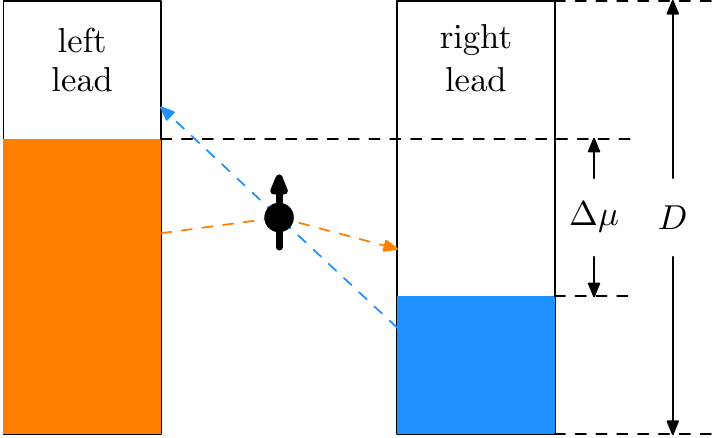}}
  \caption{(Color Online) Scheme of the spin-fermion model. The two
    leads with bandwidth $D$ are kept with a chemical potential
    difference $\Delta\mu$ at zero temperature. Electrons in the leads
    can jump into another lead via and scattered by the spin.}
  \label{fig:01}
\end{figure}

The system-bath coupling Hamiltonian \(H_{SB}\) is taken to be
\begin{equation}
  H_{SB}=\sum_{\alpha\beta;kq}V_{\alpha\beta}\sigma_zc^{\dag}_{\alpha k}c_{\beta q},
\end{equation}
where \(\alpha,\beta\) are the bath indices. With such a system-bath
coupling the momentum dependence of the scattering potential is
neglected
\cite{nozieres1969-singularities,ng1995-x,ng1996-fermi,segal2007-nonequilibrium,segal2010-numerically,chen2019-dissipative}. In
particular, only interbath system-bath coupling is under
consideration. Here we introduce a control parameter \(\lambda\) of
system-bath coupling strength for which
\(\rho V_{\alpha\beta}=\lambda(1-\delta_{\alpha\beta})\) , where
\(\rho\) is the density of states of each lead and the factor
$(1-\delta_{\alpha\beta})$ ensures only interbath
coupling. Fig. \ref{fig:01} gives a schematic representation of the
model.

In this article, we employ an iterative numerically exact influence
functional path method to investigate the effects of sweep velocity
\(v\), system-bath coupling strength \(\lambda\), and bath chemical
potential \(\Delta\mu\) on LZ transitions. This method is
nonperturbative and allows us to include nonadiabatic and
non-Markovian effects, and it is also well suited for real-time
dynamics simulations of quantum dots. It was first proposed by Makarov
and Makri for the time-independent spin-boson model
\cite{makarov1994-path,makri1995-numerical}. Later it was applied to
the monochromatically driven spin-boson model
\cite{makarov1995-control,makarov1995-stochastic,makri1997-universal,makri1997-stabilization}.
It was also adopted to investigate LZ transitions in the spin-boson
model \cite{nalbach2009-landau,arceci2017-dissipative}.

Segal \emph{et al.} generalized this method to the time-independent
spin-fermion model by adopting a discretized scheme for tracing out
the bath
\cite{segal2010-numerically,segal2011-nonequilibrium,simine2013-path,segal2013-qubit,agarwalla2017-anderson}. Chen
and Xu applied this scheme to study the monochromatically driven
spin-fermion model and gave a comparison between the path integral
method and the Floquet master equation
\cite{chen2019-dissipative}. This scheme is also adopted in this
article. The basic procedure of the path integral method is as
follows.

The evolution of total density matrix $\rho(t)$ is given by
\begin{equation}
\rho(t)=U(t)\rho(0)U^{\dag}(t),
\end{equation}
where
\begin{equation}
  U(t)=\mathrm{T}\exp[-i\int_0^tH(\tau)\dd{\tau}]=
  \prod_{t_i=0}^te^{-iH(t_i)\delta t}
\end{equation}
with $\mathrm{T}$ being the chronological ordering symbol. Here the
product is understood as that the limit is taken over all
infinitesimal interval $\delta t$ between zero and $t$ arranged from
right to left in order of increasing time $t_i$. Employing finite
$\delta t$ approximates the evolution operator $U(t)$ into a product
of finite $N$ exponentials for which $U(t)\approx\prod_{i} T_i$, where
$T_i=e^{-iH(t_i)\delta t}$. Now introduce the reduced density matrix
of the system $\rho_S=\Tr_B\rho$ by tracing $\rho$ over the bath
degrees of freedom, which is now written as
\begin{equation}
\rho_S(s'',s';t)=\Tr_B[\mel{s''}{T_N\cdots T_1\rho(0)T_1^{\dag}\cdots T_N^{\dag}}{s'}].
\end{equation}
Inserting the identity operator $\int\dd{s}\ketbra{s}{s}$ between
every two $T$ and relabeling $s'',s'$ as $s_N^+,s_N^-$ gives
\begin{equation}
\begin{split}
  \rho_S(s_N^+,s_N^-;t)=&
  \int\dd{s_0}^+\cdots\dd{s_{N-1}^+}\int\dd{s_0^-}\cdots\dd{s_{N-1}^-}\\
  &\times\Tr_B[\mel{s_N^+}{T_N}{s_{N-1}^+}\cdots\\
  &\qquad\qquad\rho(0)\cdots\mel{s_{N-1}^-}{T_N^{\dag}}{s_N^-}].
\end{split}
\end{equation}
The integrand in the above expression is referred as the ``influence
functional'' \cite{segal2010-numerically} (IF) which we denote by
$I(s_0^{\pm},\ldots,s_N^{\pm})$. The nonlocal correlations in the IF
decay exponentially under certain conditions \cite{makarov1994-path},
which enables a controlled truncation of the IF. Note that for the
spin-fermion system at zero temperature used in this article, the
exponential decay is guaranteed by finite $\Delta\mu$
\cite{segal2010-numerically,weiss2008-iterative}. Therefore the IF can
be truncated beyond a memory time $\tau_c=N_s\delta t$ with $N_s$
being a positive integer and the IF can be written approximately as
\cite{makarov1994-path,makri1995-numerical,makri1999-iterative,segal2010-numerically,segal2011-nonequilibrium}
\begin{equation}
\label{eq:IF}
\begin{split}
  I(s_0^{\pm},\ldots,s_N^{\pm})\approx& I(s_0^{\pm},\ldots,s_{N_s}^{\pm})
  I_s(s_1^{\pm},\ldots,s_{N_s+1}^{\pm})\\
  &\quad\times\cdots I_s(s_{N-N_s},\ldots,s_N^{\pm}),
\end{split}
\end{equation}
where
\begin{equation}
I_s(s_k^{\pm},\ldots,s_{k+N_s}^{\pm})=\frac{I(s_k^{\pm},\ldots,s_{k+N_s}^{\pm})}{I(s_k^{\pm},\ldots,s_{k+N_s-1}^{\pm})}.
\end{equation}
In order to integrate Eq. \eqref{eq:IF} iteratively we can define a
multiple time reduced density matrix
$\tilde{\rho}_S(s_k^{\pm},\ldots,s_{k+N_s-1}^{\pm})$ with initial
values $\tilde{\rho}_S(s_0^{\pm},\ldots,s_{N_s-1}^{\pm})=1$. Its first
evolution step is given by
\begin{equation}
  \tilde{\rho}_S(s_1^{\pm},\ldots,s_{N_s}^{\pm})=\int\dd{s_0^{\pm}}
  I(s_0^{\pm},\ldots,s_{N_s}^{\pm}),
\end{equation}
and the latter evolution step is given by
\begin{equation}
\begin{split}
  \tilde{\rho}_S(s_{k+1}^{\pm},\ldots,s_{k+N_s}^{\pm})=&\int\dd{s_k^{\pm}}
  \tilde{\rho}_S(s_k^{\pm},\ldots,s_{k+N_s-1}^{\pm})\\
  &\quad\times I_s(s_k^{\pm},\ldots,s_{k+N_s}^{\pm}).
\end{split}
\end{equation}
Finally the time-local ($t_k=k\delta t$) reduced density matrix is
obtained by
\begin{equation}
  \rho_S(t_k)=\int\dd{s_{k-1}^{\pm}}\cdots\dd{s_{k-N_s+1}^{\pm}}
  \tilde{\rho}_S(s_{k-N_s+1}^{\pm},\ldots,s_k^{\pm}).
\end{equation}

It can be seen that we need to keep track of a $2N_s$ rank ``tensor''
$\tilde{\rho}_S$ and a $2(N_s+1)$ rank ``tensor'' $I_s$. If the size
of the system Hilbert space is $M$, then a space with size
proportional to $M^{2N_s}$ is needed to store $\tilde{\rho}_S$ and a
space with size proportional to $M^{2(N_s+1)}$ is needed to store
$I_s$. The space size increases dramatically with increasing $M$ and
$N_s$, which limits the value of time step $\delta t$, the length of
$\tau_c$ and the size of the system $M$ in a practical
calculation. However, because the method is iterative in time it is
easy to deal with a time-dependent Hamiltonian.

In principle, the final results of the time-independent model can be
extrapolated to the $\delta t\to0$ limit and the error brought by
finite $\delta t$ is then eliminated
\cite{segal2010-numerically,weiss2008-iterative}. However, in
time-dependent driving case with different $\delta t$ the driving
field is sampled in different time grids, which would bring extra
error in extrapolation. In addition, $\delta t$ can not be arbitrary
small with a fixed $\tau_c$ since we must ensure that $N_s$ is not too
large. Therefore, as in Ref. \cite{chen2019-dissipative}, the
extrapolation is not employed in this article.

\section{Results and Discussions}
\label{sec:result}

In the pure LZ problem, at initial time \(t=-\infty\) the system is
prepared in one diabatic state, and one seeks the probability \(P_0\)
of the system to end up in another at \(t=+\infty\). If the system is
initially prepared in the up state $\ket{\uparrow}$, which corresponds
to the ground state, then \(P_0\) gives the probability of the system
to end up in the down state $\ket{\downarrow}$, which now also
corresponds to the ground state. In other words, the LZ probability
\(P_0\) gives the final probability of the system to stay in the
ground state. Similarly, if the system is initially prepared in the
down state $\ket{\downarrow}$ then \(P_0\) gives the final probability
to stay in the excited state. In summary, $P_0$ is defined as
\begin{equation}
  P_0=\abs{\mel{\uparrow}{U_S(\infty,-\infty)}{\downarrow}}^2=
  \abs{\mel{\downarrow}{U_S(\infty,-\infty)}{\uparrow}}^2,
\end{equation}
where $U_S$ is the system evolution operator 
\begin{equation}
  U_S(\infty,-\infty)=\mathrm{T}\exp[-i\int_{-\infty}^{\infty}H_S(\tau)\dd{\tau}].
\end{equation}

The exact solution for \(P_0\) is given by
\cite{landau1932-on,zener1932-non,stueckelberg1932-theorie,majorana1932-atomi}
\begin{equation}
  P_0=1-\exp(-\frac{\pi\Delta_0^2}{2v}).
\end{equation}
This solution is symmetric for both diabatic states for which whether
the system is prepared in the up or down state would not affect the
probability of it transiting to another diabatic state.

When the system is coupled to the environment, this symmetry is broken
for which the probability of the system staying in the ground or
excited state becomes different. For convenience, we denote the LZ
probability of the system to stay in the ground state (corresponding
to the transition from up to down state) by \(P_1\), for which
\begin{equation}
P_1=\abs{\mel{\downarrow}{U(\infty,-\infty)}{\uparrow}}^2,
\end{equation}
and the LZ probability of the system to stay in the excited state
(corresponding to the transition from down to up state) by \(P_2\),
for which
\begin{equation}
P_2=\abs{\mel{\uparrow}{U(\infty,-\infty)}{\downarrow}}^2.
\end{equation}
Here $U$ denotes the total evolution operator for which
\begin{equation}
  U(\infty,-\infty)=\mathrm{T}\exp[-i\int_{-\infty}^{\infty}H(\tau)\dd{\tau}].
\end{equation}
In this section the simulation results for \(P_1\) and \(P_2\) are
given respectively. In all figures, the simulation results are
presented by dots and lines are guides for the eye.

\subsection{Results of \(P_1\)}
\label{sec:P1}

Let us first consider the case where the system is initially prepared
in the up (ground) state at \(t=-\infty\).

\begin{figure}[htbp]
  \centerline{\includegraphics[]{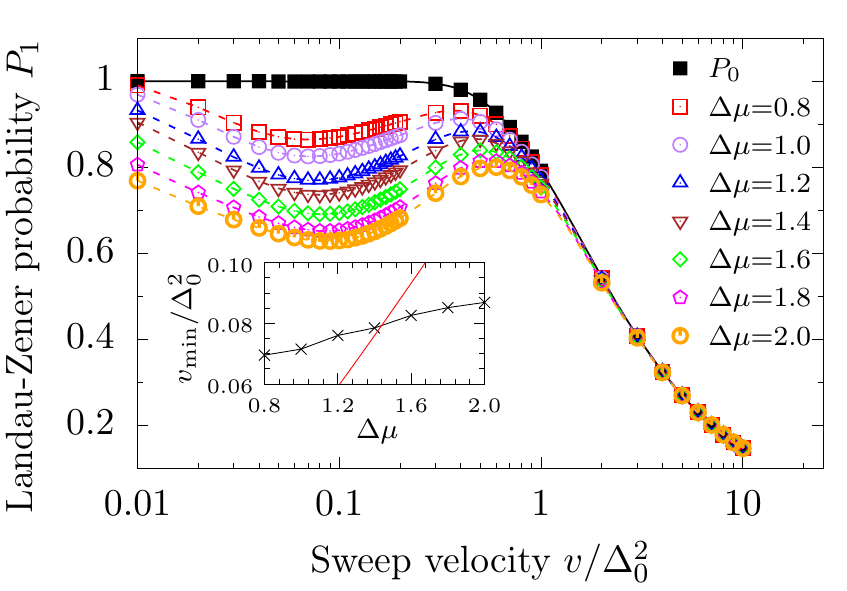}}
  \caption{(Color Online) The LZ probability $P_1$ versus $v$ for weak
    system-bath coupling strength $\lambda=0.04$ and various
    $\Delta\mu$. Inset: $v_{\mathrm{min}}$ versus $\Delta\mu$ with
    same $\lambda$ as in the main figure. The simulation results are
    shown by black dots, and results by the phenomenological model
    [Eq. \eqref{eq:time-scale}] are shown by the red line.}
  \label{fig:02}
\end{figure}

Figure \ref{fig:02} shows the LZ probability \(P_1\) versus sweep
velocity \(v\) for weak coupling \(\lambda=0.04\) and various
\(\Delta\mu\). It can be seen that a large velocity regime
(\(v\gg\Delta_0^2\)) can be distinguished from a small velocity regime
(\(v\ll\Delta_0^2\)). The result shown here is similar to that in the
spin-boson model for various temperatures \cite{nalbach2009-landau}
for which larger \(\Delta\mu\), which can act as a temperature like
dephasing contributor \cite{segal2007-nonequilibrium}, suppresses the
LZ transition more strongly. In the large-velocity regime
(\(v\gg\Delta_0^2\)), \(P_1\) coincides with \(P_0\) for which little
influence of the environment is shown. In the regime where
\(v<\Delta_0^2\), besides an overall all decrease of \(P_1\) with
increasing \(\Delta\mu\), a nonmonotonic dependence of \(P_1\) on
\(v\) is shown: with fixed \(\Delta\mu\) and decreasing \(v\), \(P_1\)
first shows a maximum at \(v_{\mathrm{max}}\), which is smaller than
but close to \(\Delta_0^2\), then a minimum at velocity
\(v_{\mathrm{min}}\), and finally an increase.

\begin{figure}[htbp]
  \centerline{\includegraphics[]{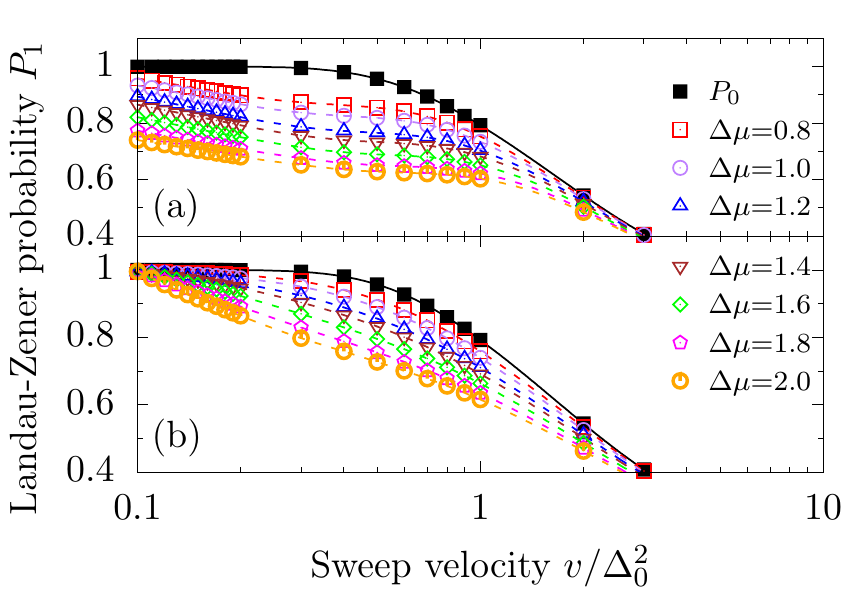}}
  \caption{(Color Online) Same as Fig. \ref{fig:02} for larger
    system-bath coupling strengths (a) $\lambda=0.10$ and (b)
    $\lambda=0.20$.}
\label{fig:03}
\end{figure}

In the spin-boson model, this nonmonotonicity can not be described by
perturbative approaches but can be explained by a simple
phenomenological model \cite{nalbach2009-landau}. Here we give a
review on such a phenomenological model. The bath is assumed to mainly
induce relaxation. At the initial time the system is prepared in the
ground state; therefore only absorption can occur if an excitation
with energy
\begin{equation}
  \Delta_t=\sqrt{(vt)^2+\Delta_0^2}
\end{equation}
exists in the bath spectrum and thermally populated. The quantity
\(\Delta_t\) varies with time and if it is larger than a threshold
energy \(\Delta_c\) then relaxation would stop. In other words,
relaxation can only occur within a time window from
\(-\frac{1}{2}t_r\) to \(\frac{1}{2}t_r\), where
\begin{equation}
  t_r=\frac{2}{v}\sqrt{\Delta_c^2-\Delta_0^2}.
  \label{eq:tr}
\end{equation}
The threshold energy \(\Delta_c\) is taken to be the smaller of the
temperature \(T\) and the bath cutoff frequency \(\omega_c\). Let
\(\tau_r\) denote the system relaxation time; then \(\tau_r\) must be
shorter than \(t_r\) for relaxation processes to contribute.

When the sweep velocity \(v\) is large, \(t_r\) is small for which
\(t_r\ll\tau_r\); therefore relaxation can not occur and the bath has
little influence on the LZ transition for which \(P_1\) coincides with
\(P_0\). In the opposite limit where \(v\to0\) for which
\(t_r\gg\tau_r\), the system will get full relaxation at any time
according to momentary Hamiltonian. Since relaxation stops at the
threshold energy, the system would be adjusted according to
\(\Delta_c\) and \(T\). For small but finite sweep velocity,
equilibration is retarded for which the system is relaxed according to
the past momentary Hamiltonian. The system is then assumed to be
equilibrated toward a time-averaged energy splitting
\begin{equation}
  \bar{\Delta}_r=\frac{1}{t_r}\int_{-t_r/2}^{t_r/2}\Delta_t\dd{t},
\end{equation}
leading to
\(P_1(v_{\mathrm{min}})=\frac{1}{2}[1+\tanh(\bar{\Delta}_r/2T)]\).

According to the discussion, large \(t_r\) or small \(t_r\) would
weaken the suppression of the LZ transition. Thus it is assumed that
relaxation will maximally suppress the LZ transition when \(t_r\) and
\(\tau_r\) coincide,
\begin{equation}
  t_r(v_{\mathrm{min}})=\tau_r,
  \label{eq:time-scale}
\end{equation} 
which leads to a minimum of \(P_1\) at \(v_{\mathrm{min}}\). If only
single-phonon absorption is considered within resonance
(\(\abs{t}\le\frac{t_r}{2}\)), then the system relaxation time
\(\tau_r\) can estimated by the golden rule with time-averaged energy
splitting \(\bar{\Delta}_r\) for which
\begin{equation}
  \tau_r^{-1}=\pi\alpha\frac{\Delta_0^2}{\bar{\Delta}_r}
  \exp(-\bar{\Delta}_r/\omega_c)n(\bar{\Delta}_r),
\label{eq:boson-tau-r}
\end{equation} 
where \(\alpha\) is the system-bath coupling strength and
\(n(\bar{\Delta}_r)\) is the Bose-Einstein distribution function. A
revisit of the golden rule used in the spin-boson model is given in
the Appendix. Comparing Eq. \eqref{eq:tr} and \eqref{eq:boson-tau-r}
gives the position of \(v_{\mathrm{min}}\).

Now apply this phenomenological model to our spin-fermion model. Since
the system is prepared in the ground state, only absorption can occur
if an electron jumps from the left lead to an unoccupied state with
lower energy in the right lead. The energy difference of the electron
before and after the jump should be \(\Delta_t\). The largest energy
change by the jump is \(\Delta\mu\) (when an electron at the Fermi
level in the left lead jump to the Fermi level of the right lead);
thus \(\Delta_c=\Delta\mu\).

When the sweep velocity \(v\) is large for which \(t_r\ll\tau_r\), the
relaxation can not occur and thus \(P_1\) coincides with \(P_0\).
When the sweep velocity \(v\) is small for which \(t_r\gg\tau_r\), the
system would be fully relaxed according to \(\Delta_c=\Delta\mu\) at
zero temperature.  In the spin-fermion model, the full relaxation of
the system is determined not by the temperature but by the chemical
potential difference \(\Delta\mu\). According to
Ref. \cite{segal2007-nonequilibrium}, it has no simple analytical
formula for the system polarization \(\expval{\sigma_z}\), but it is
known that \(\expval{\sigma_z}\) manifests a transition from a fully
polarized system, where the system is in the ground state, to an
unpolarized system, where \(\expval{\sigma_z}=0\) as \(\Delta\mu\)
increases. Since $\Delta_c=\Delta\mu$, which is equivalent to say
$\Delta\mu$ is not large, after fully relaxation the system would be
adjusted to the ground state.  Therefore \(P_1=1\) when \(v\to0\), and
this can be seen more clearly from Fig. \ref{fig:03}(b) where a larger
coupling strength \(\lambda\) accelerates relaxation processes.

\begin{figure}[htbp]
  \centerline{\includegraphics[]{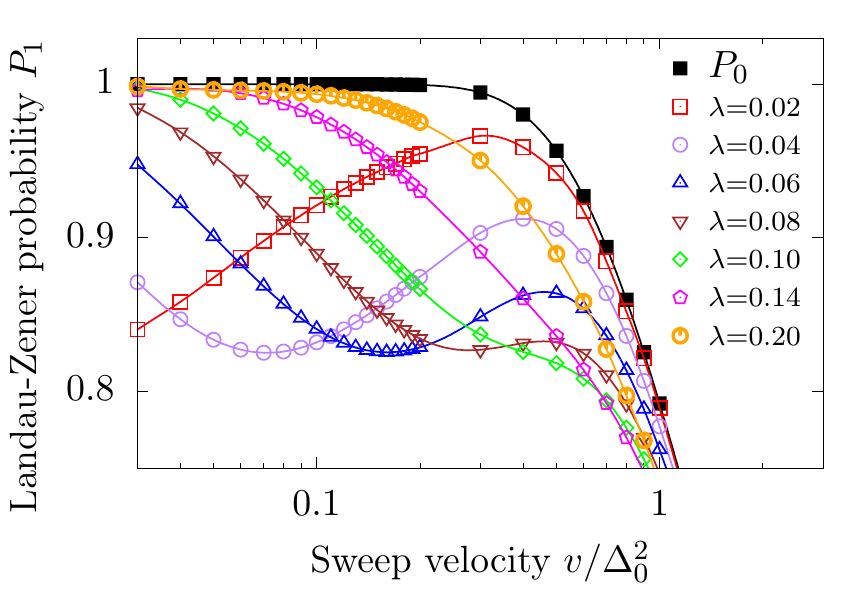}}
  \caption{(Color Online) The LZ probability $P_1$ versus $v$ for
    $\Delta\mu=1.0$ and various $\lambda$.}
  \label{fig:04}
\end{figure}

When \(t_r\) matches \(\tau_r\), the relaxation maximally suppresses
the LZ transition, which gives a minimum at
\(v_{\mathrm{min}}\). Since the value of \(P_1(v_{\mathrm{min}})\)
must be smaller than \(P_1(v\to0)=1\), as \(v\to0\) the probability
\(P_1\) must increase again with decreasing \(v\). Therefore \(P_1\)
shows a maximum at \(v_{\mathrm{max}}\), and minimum at
\(v_{\mathrm{min}}\), and then an increase with decreasing \(v\), as
shown in Fig. \ref{fig:02}. Basically, the mechanism for nonmonotonic
dependence on \(v\) in the spin-fermion model is same as that in the
spin-boson model. If the system ends up in an unpolarized state with
$\expval{\sigma_z}=0$ then we have \(P_1=\frac{1}{2}\). This is the
reason why \(P_1(v_{\mathrm{min}})\) deviates from \(P_0\) and is
closer to \(\frac{1}{2}\) for larger \(\Delta\mu\), or in other words,
larger \(\Delta\mu\) would suppress the LZ transition more strongly.

\begin{figure}[htbp]
  \centerline{\includegraphics[]{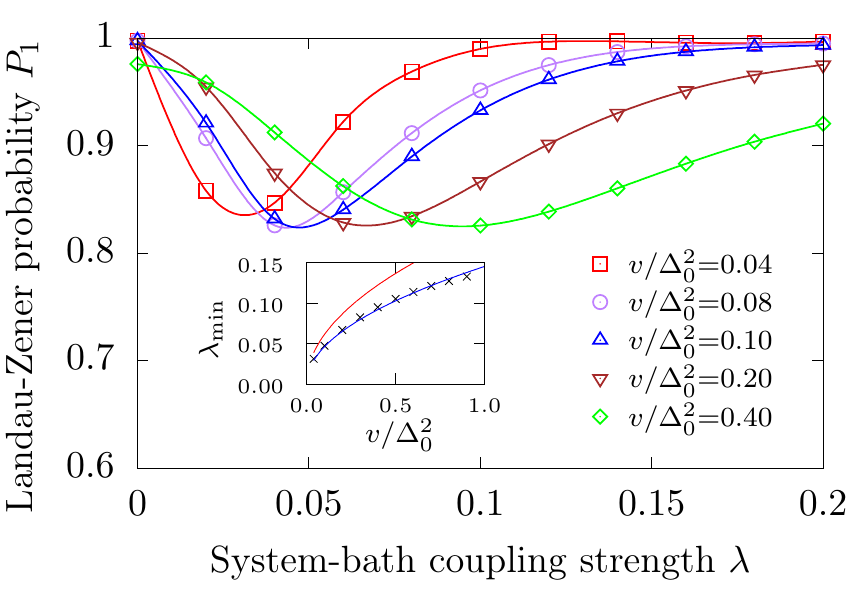}}
  \caption{(Color Online) The LZ probability $P_1$ versus $\lambda$
    for $\Delta\mu=1.0$ and various $v$. Inset:
    $\lambda_{\mathrm{min}}$ versus $v$ with same $\Delta\mu$ in the
    main figure. The simulation data are shown by black dots, results
    by the phenomenological model are shown by the red line, and the
    blue line represents a fitting function $y=0.145x^{\frac{1}{2}}$.}
  \label{fig:05}
\end{figure}

For a fixed time and weak coupling, the decay rate out of the ground
state can be estimated by the golden rule if only a single electron
jump is considered. After summing up all possible jumps whose energy
difference is \(\Delta_t\), we obtain the decay rate as (see the
Appendix)
\begin{equation}
  \tau^{-1}=2\pi\lambda^2\frac{\Delta_0^2}{\Delta_t^2}(\Delta\mu-\Delta_t).
  \label{eq:decay-rate-ground}
\end{equation}
In the spin-boson model, \(\tau_r^{-1}\) is archived via simply
substituting \(\Delta_t\) by \(\bar{\Delta}_r\) in the expression of
\(\tau^{-1}\). However, in the spin-fermion model, due to the inverse
quadratic dependence on \(\Delta_t\) of \(\tau^{-1}\), it would be
more appropriate to estimate \(\tau_r^{-1}\) by the time-averaged
decay rate
\begin{equation}
  \begin{split}
    \tau_r^{-1}&=\frac{1}{t_r}\int_{-t_r/2}^{t_r/2}2\pi\lambda^2
    \frac{\Delta_0^2}{\Delta_t^2}(\Delta\mu-\Delta_t)\dd{t}\\
    &=\frac{4\pi}{vt_r}\lambda^2\Delta_0^2
    \qty[\frac{\Delta\mu}{\Delta_0}\atan(\frac{vt_r}{2\Delta_0})
    -\mathrm{asinh}(\frac{vt_r}{2\Delta_0})].\\
  \end{split}
\label{eq:relaxation-ground}
\end{equation}
This formula predicts that \(v_{\mathrm{min}}\) increases almost
linearly with increasing \(\Delta\mu\) when
\(\Delta_c\gg\Delta_0\). The positions of $v_{\mathrm{min}}$ versus
$\Delta\mu$ for $\lambda=0.04$ is shown in the inset of
Fig. \ref{fig:02}. It can be seen that $v_{\mathrm{min}}$ roughly
shows a linearly dependence on $\Delta\mu$. However, employing
Eq. \eqref{eq:time-scale} we have
$v_{\mathrm{min}}\approx0.043\Delta_0^2$ for $\Delta\mu=1.0$ and
$v_{\mathrm{min}}\approx0.127\Delta_0^2$ for $\Delta\mu=2.0$. This
result only qualitatively agrees with what is shown in
Fig. \ref{fig:02} where $v_{\mathrm{min}}\approx0.072\Delta_0^2$ for
$\Delta\mu=1.0$ and $v_{\mathrm{min}}\approx0.087\Delta_0^2$ for
$\Delta\mu=2.0$.

Figure \ref{fig:03} shows the LZ transition probability \(P_1\) versus
sweep velocity \(v\) for larger $\lambda$ and various
\(\Delta\mu\). As seen from Eq. \eqref{eq:relaxation-ground}, the
relaxation rate is proportional to \(\lambda^2\) for which increasing
\(\lambda\) enhances relaxation and decreases \(\tau_r\)
accordingly. For \(\lambda=0.1\) [Fig. \ref{fig:03}(a)] the minimum
disappears and only a shoulder remains. For \(\lambda=0.2\)
[Fig. \ref{fig:03}(b)], only a monotonic growth of \(P_1\) with
decreasing \(v\) remains, and the relaxation processes are greatly
accelerated for which at \(v/\Delta_0^2=0.1\) the LZ transition
probabilities \(P_1\) already reduce to 1.

Figure \ref{fig:04} shows the LZ transition probability \(P_1\) versus
\(v\) for \(\Delta\mu=1.0\) and various \(\lambda\). For small
\(\lambda=0.04,0.06,0.08\), the minimum \(P_1(v_{\mathrm{min}})\) can
be still observed and \(v_{\mathrm{min}}\) shifts to larger velocity
for increasing \(\lambda\). The minimum disappears when
\(\lambda\ge0.10\) and in this case \(P_1\) goes to 1 for small
\(v\). Due to these features, the lines of \(P_1\) in
Fig. \ref{fig:04} have cross points, which means there is a
nonmonotonic dependence of \(P_1\) on \(\lambda\). This can be seen
more clearly in Fig. \ref{fig:05}.

Figure \ref{fig:05} shows the dependence of \(P_1\) on \(\lambda\) for
\(\Delta\mu=1.0\) and various \(v\). It can be seen that \(P_1\) shows
a minimum at \(\lambda_{\mathrm{min}}\) for a fixed \(v\). Since
\(t_r^{-1}\propto v\) and \(\tau_r^{-1}\propto\lambda^2\), we have
$\lambda_{\mathrm{min}}^2\propto v$ which means there is a simple
quadratic dependence between $v$ and $\lambda_{\mathrm{min}}$: when
\(v\) is scaled by a factor of \(a\) then \(\lambda_{\mathrm{min}}\)
would be scaled by a factor of \(\sqrt{a}\). This conclusion agrees
with the results shown in the inset of the figure where a inverse
quadratic fitting is shown.

Equation \eqref{eq:relaxation-ground} gives a simple description of
the effect of \(\lambda\) on \(\tau_r^{-1}\), while the effect of
\(\Delta\mu\) is much more complex. It is because \(\Delta\mu\) plays
a role as both temperature \(T\) and bath cutoff frequency
\(\omega_c\) in the spin-boson model, which makes its effect on the
relaxation complex, while larger \(\lambda\) simply enhances the
relaxation. In the spin-boson model, it is already mentioned that by
the phenomenological picture the behavior of the crossover temperature
can be only roughly described with absolute values off by a factor of
3 \cite{nalbach2009-landau}. This may be the reason why the results of
Eq. \eqref{eq:relaxation-ground} always have some deviations from
simulation data since the effect of $\omega_c$ can not be removed from
$\Delta\mu$.

\subsection{Results of \(P_2\)}
\label{sec:P2}

Now let us turn to \(P_2\) which stands for the LZ transition
probability from down to up state, where the system is prepared in the
excited state at \(t=-\infty\).

\begin{figure}[htbp]
  \centerline{\includegraphics[]{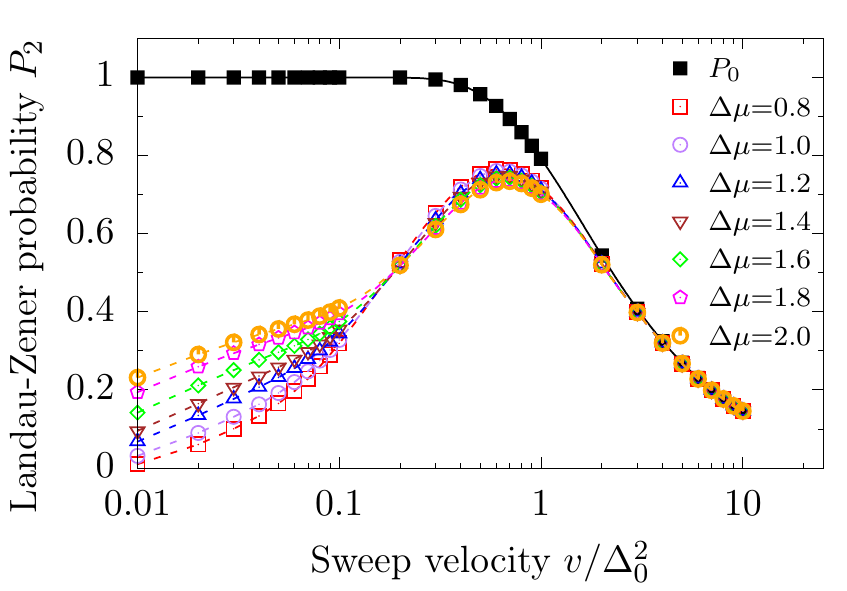}}
  \caption{(Color Online) The LZ probability $P_2$ versus $v$ for weak
    system-bath coupling strength $\lambda=0.04$ and various
    $\Delta\mu$.}
  \label{fig:06}
\end{figure}

\begin{figure}[htbp]
  \centerline{\includegraphics[]{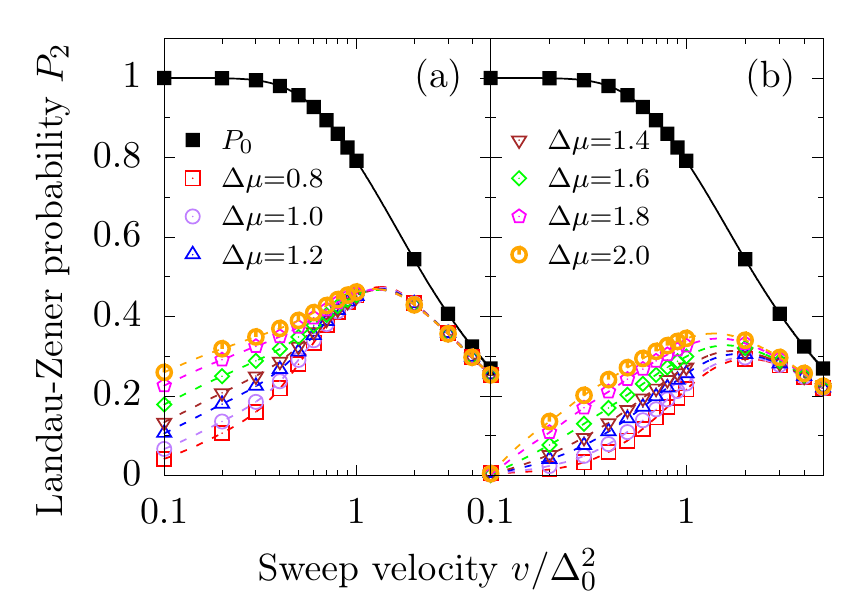}}
  \caption{(Color Online) The same as Fig. \ref{fig:06} for larger
    system-bath coupling strengths (a) $\lambda=0.10$ and (b)
    $\lambda=0.20$.}
  \label{fig:07}
\end{figure}

Figure \ref{fig:06} shows the LZ probability \(P_2\) versus \(v\) for
weak coupling \(\lambda=0.04\) and various \(\Delta\mu\). In large
sweep velocity regime (\(v\gg\Delta_0^2\)), since \(t_r\) is too short
for relaxation \(P_2\) coincides with \(P_0\), just like \(P_1\). In
the small sweep velocity regime (\(v\ll\Delta_0^2\)), there is great
difference between the behaviors of \(P_2\) and \(P_1\). For small
\(v\), \(P_1\) is close to \(P_0\), i.e., close to 1, while \(P_2\)
is far away from \(P_0\) and close to zero. With fixed \(\Delta\mu\)
and decreasing \(v\), \(P_2\) shows a maximum at \(v_{\mathrm{max}}\),
which is smaller than but close to \(\Delta_0^2\), then decreases all
along and no minimum is shown.

Although behaviors of \(P_1\) and \(P_2\) differ greatly for small
\(v\), they are due to the same relaxation mechanism. For \(P_1\), we
are seeking the probability of the system to stay in the ground state,
while for \(P_2\), the probability of the system to stay in the
excited state is desired. However, when \(v\) is small a full
relaxation would lead the system towards the ground state, and this
makes \(P_1\) go to 1 and \(P_2\) to zero.

There is another difference between the behaviors of \(P_1\) and
\(P_2\) in the small-\(v\) regime: for \(P_1\), larger \(\Delta\mu\)
suppresses the LZ transition probability more strongly, while for
\(P_2\), larger \(\Delta\mu\) increases the LZ transition probability
instead. This is because, as mentioned earlier, larger \(\Delta\mu\)
would relax the system toward an unpolarized state, where
\(\expval{\sigma_z}=0\), which makes \(P_2\) closer to
\(\frac{1}{2}\). Around \(v_{\mathrm{max}}\), the situation is in
another way around for which larger \(\Delta\mu\) decreases
\(P_2\). The underlying reason is the same: larger \(\Delta\mu\) makes
the system go toward an unpolarized state, i.e., makes \(P_2\) closer
to \(\frac{1}{2}\).

Since the system is initially prepared in the excited state, only
emission can occur within resonance (\(\abs{t}\le\frac{t_r}{2}\)). In
the emission process, there are two kinds jumping: an electron in the
left lead jumps to an unoccupied state with higher energy in the right
lead, and an electron in the right lead jumps to an unoccupied state
in the left lead with higher energy. The energy difference by the jump
should be \(\Delta_t\). If only a single electron jump is considered,
the golden rule formula for the decay rate out of the excited state is
given in the Appendix. Basically, the golden rule states that the
decay rate \(\tau^{-1}\) has a quadratic dependence on \(\lambda\) as
in the absorption process, while its dependence on \(\Delta\mu\) is of
more complexity.

\begin{figure}[htbp]
  \centerline{\includegraphics[]{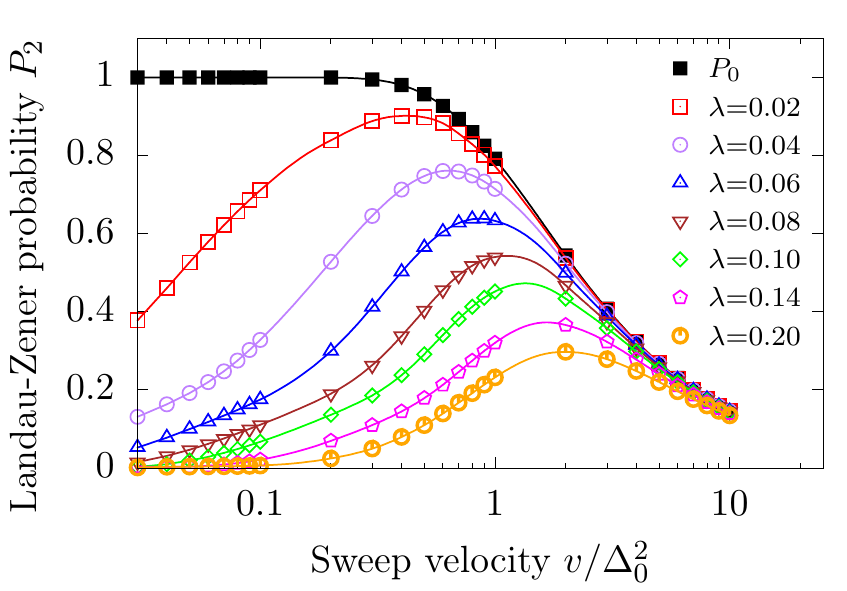}}
  \caption{(Color Online) The LZ probability $P_2$ versus $v$ for
    $\Delta\mu=1.0$ and various $\lambda$.}
  \label{fig:08}
\end{figure}

\begin{figure}[htbp]
  \centerline{\includegraphics[]{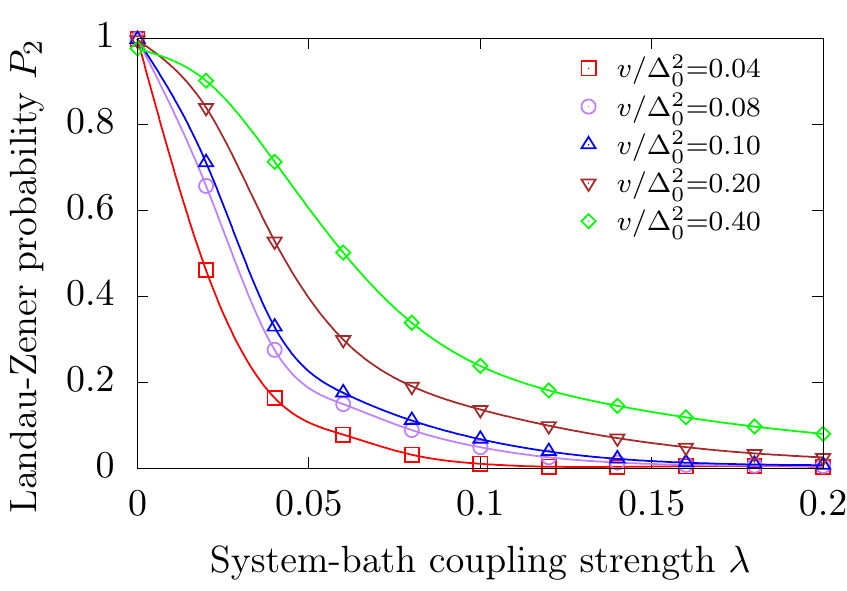}}
  \caption{(Color Online) The LZ probability $P_2$ versus $\lambda$
    for $\Delta\mu=1.0$ and various $v$.}
  \label{fig:09}
\end{figure}

Figure \ref{fig:07} shows the same content as Fig. \ref{fig:06} for
larger system-bath coupling strengths. Since the relaxation rate is
proportional to \(\lambda^2\), larger \(\lambda\) accelerates
relaxation and decreases \(\tau_r\). It can be seen that with
\(\Delta\mu=2.0\) when \(\lambda=0.04\) (Fig. \ref{fig:06}) \(P_2\) is
close to 0.2 at \(v=0.01\Delta_0^2\), while when \(\lambda=0.1\)
[Fig. \ref{fig:07}(b)] \(P_2\) is already close to 0.2 at
\(v=0.1\Delta_0^2\). For \(\lambda=0.2\) [Fig. \ref{fig:07}(b)],
\(P_2\) already reduces to zero at \(v=0.1\Delta_0^2\).

Figure \ref{fig:08} shows the LZ transition probability \(P_2\) versus
\(v\) for \(\Delta\mu=1.0\) and various \(\lambda\). Unlike \(P_1\)
shown in Fig. \ref{fig:04}, lines of \(P_2\) in Fig. \ref{fig:08} have
no cross points. This is because there is no \(v_{\mathrm{min}}\) in
\(P_2\). In addition, due to the same reason the dependence of
\(\lambda\) of \(P_2\) becomes monotonic. This can be seen from
Fig. \ref{fig:09} which shows \(P_2\) versus \(\lambda\) for
\(\Delta\mu=1.0\) and various \(v\). It can be seen that larger
\(\lambda\) accelerates the relaxation and makes \(P_2\) go to zero,
and smaller \(v\) also enhances the relaxation.

\section{Conclusions}
\label{sec:conclusion} 

We have studied LZ transitions in a fermionic environment where a TLS
is coupled to two metallic leads kept with different chemical
potential at zero temperature. The dynamics of the system is simulated
by an iterative numerically exact influence functional path integral
method which allows us to include nonadiabatic and non-Markovian
effects.

The LZ transition probability is the probability of the system staying
in the ground (excited) state \(P_1\) (\(P_2\)) after an energy
sweep. In the pure LZ problem, the two kinds of probability are
symmetric; i.e., they are the same no matter whether the system is
initially prepared in the ground or excited state.

The symmetry no longer exists after taking the effect of the
environment into consideration. In the large sweep velocity regime,
\(v\gg\Delta_0^2\), since the resonance time \(t_r\) is much shorter
than the system relaxation time \(\tau_r\), the bath has little
influence on the transition; thus both \(P_1\) and \(P_2\) coincide
with the pure LZ transition probability \(P_0\). In the small sweep
velocity regime, \(v\ll\Delta_0^2\), the system is fully relaxed to
the ground state, which makes \(P_1\) go to 1 and \(P_2\) to
zero. This is the reason why the symmetry no longer exists. Due to the
same reason, \(P_1\) shows a minimum at \(v_{\mathrm{min}}\) and
\(\lambda_{\mathrm{min}}\), while \(P_2\) shows no minimum.

According to the phenomenological model, the existence of
\(v_{\mathrm{min}}\) of \(P_1\) is understood as a nontrivial
competition between relaxation and LZ driving, where the LZ transition
is maximally suppressed when the resonance time \(t_r\) and the system
relaxation time \(\tau_r\) coincide. The system relaxation time
\(\tau_r\) can be estimated by the golden rule, which states that
\(\tau_r^{-1}\) has a quadratic dependence on the system-bath coupling
strength \(\lambda\). This statement agrees with our results, which
indicates that the effect of \(\lambda\) is fairly simple.

The effect of $\Delta\mu$ on the dissipation is of more complexity. If
we treat $\Delta\mu$ as the temperature, then results similar to those
shown in Figs. \ref{fig:02}, \ref{fig:03}, and \ref{fig:04} can be
found in the LZ transitions in the spin-boson model
\cite{nalbach2009-landau}. This is because $\Delta\mu$ can act as a
temperature like dephasing contributor. However, it is not really the
temperature and the temperature also has its own effect on the LZ
transitions. At zero temperature, the system manifests a transition
from the ground state to an unpolarized system as $\Delta\mu$
increases. This means that when the value of $\Delta_t$ can compare
with $\Delta\mu$, then the system would be relaxed to the ground
state.  Meanwhile in the spin-boson model, the system would not be
relaxed to the ground state at finite temperature, and the bath shows
no effect on the LZ probability when the system and bath are
diagonally coupled at zero temperature \cite{wubs2006-gauging}.

In addition, $\Delta\mu$ also plays the role of the cutoff frequency
$\omega_c$ of the spin-boson model, where the effect of $\omega_c$ can
be also only qualitatively described by the phenomenological
model. The dual role of $\Delta\mu$ as both the temperature $T$ and
the cutoff frequency $\omega_c$ in the spin-boson model makes its
effect on the dissipation complex. This may be the reason why results
by the phenomenological model always have some deviations from
simulation data since the effect of $\omega_c$ can not be removed from
$\Delta\mu$. The interplay between the external field, system-bath
coupling \(\lambda\) and chemical potential difference $\Delta\mu$
remains open for further investigations.

Despite the different effects of the bosonic and fermionic baths
discussed above, the phenomenological model also gives some universal
predictions despite what kind of environment is present: when the
sweep velocity is large, due to small relaxation time window $t_r$ the
effect of the environment can be neglected; when the sweep velocity is
small, the system would be fully relaxed according to the environment,
where the environment shows its characteristics most; in the
intermediate sweep velocity regime, the LZ probability shows a
nonmonotonic dependence on the sweep velocity and the coupling
parameter. This nonmonotonic feature may be useful for optimal control
problems and for further experiments.

\appendix
\section{Golden Rule}
\label{app:golden}

In this appendix, we shall revisit the derivation of the golden rule
formula used for the LZ transition in the spin-boson model
\cite{nalbach2009-landau}, and then following the same spirit we shall
derive the golden rule for the spin-fermion model studied in this
article.

\subsection{Spin-Boson Model}
The Hamiltonian of the spin-boson model is
\begin{equation}
  H=H_S+H_B+H_{SB},
\end{equation}
where the system and bath Hamiltonian are
\begin{equation}
  H_S=\frac{\varepsilon}{2}\sigma_z+\frac{\Delta_0}{2},\quad
  H_B=\sum_k\omega_kb_k^{\dag}b_k.
\end{equation}
Here \(\varepsilon\) and \(\Delta_0\) are level splitting and
tunneling amplitude of the TLS, respectively, and \(b_k\)
(\(b_k^{\dag}\)) are bosonic annihilation (creation) operators. For
simplicity we assume \(\varepsilon\) is positive. The system-bath
coupling is written as
\begin{equation}
  H_{SB}=\frac{\sigma_z}{2}\sum_k\lambda_k(b_k+b_k^{\dag}),
\end{equation}
and the bath influence is described by the spectral density
\begin{equation}
  J(\omega)=\sum_k\lambda_k^2\delta(\omega-\omega_k)
  =2\alpha\omega\exp(-\omega/\omega_c),
\end{equation}
where an Ohmic form with cutoff frequency \(\omega_c\) and coupling
strength \(\alpha\) is considered.

The system Hamiltonian \(H_S\) can be diagonalized by an unitary
rotation for which
\begin{equation}
  \bar{H}_S=e^{\frac{1}{2}i\theta\sigma_y}H_Se^{-\frac{1}{2}i\theta\sigma_y}
  =\frac{1}{2}\Delta_t\sigma_z
  \label{eq:rotation}
\end{equation}
when \(\tan\theta=\Delta_0/\varepsilon\), where
\(\Delta_t=\sqrt{\varepsilon^2+\Delta_0^2}\). After the rotation, the
system-bath coupling becomes
\begin{equation}
  \bar{H}_{SB}=\frac{1}{2\Delta_t}(\varepsilon\sigma_z-\Delta_0\sigma_x)
  \sum_k\lambda_k(b_k+b_k^{\dag}).
\end{equation}

The Pauli matrix \(\sigma_x\) can be written in terms of
\(\sigma_+=\frac{1}{2}(\sigma_x+i\sigma_y)\) and
\(\sigma_-=\frac{1}{2}(\sigma_x-i\sigma_y)\) for which
\(\sigma_x=\sigma_++\sigma_-\). If only a single phonon absorption is
considered in the absorption process, then in \(\bar{H}_{SB}\) only
the term \(\frac{1}{2}\frac{\Delta_0}{\Delta_t}\sum_k\sigma_+b_k\) is
relevant. Let \(\ket{1}\) denote the excited state (\(\sigma_z=+1\))
and \(\ket{2}\) denote the ground state (\(\sigma_z=-1\)), and let
\(\ket{n_k}\) denote the phonon state
\(\frac{1}{\sqrt{n_k!}}(b_k^{\dag})^{n_k}\ket{0}\), where \(\ket{0}\)
is the vacuum state. Then starting from the state \(\ket{2}\), the
probability \(p(t)\) of the system to go to the state \(\ket{1}\) at
time \(t\) is given by the golden rule formula
\cite{leggett1987-dynamics,weiss1993-quantum}
\begin{equation}
  \begin{split}
    p(t)=&\frac{1}{4}\frac{\Delta_0^2}{\Delta_t^2}\int_0^t\int_0^t
    \sum_k\rho_k\lambda_k^2e^{i(\Delta_t-\omega_k)(t_1-t_2)}\\
    &\quad\times\abs{\mel{1}{\sigma_+}{2}\mel{n_k-1}{b_k}{n_k}}^{2}
    \dd{t_1}\dd{t_2},
  \end{split}
  \label{eq:golden}
\end{equation}
where \(\rho_k\) is the Gibbs distribution function
\begin{equation}
  \rho_k=\exp(-\frac{n_k\omega_k}{T})/
  \sum_{n_k}\exp(-\frac{n_k\omega_k}{T})
\end{equation}
with temperature \(T\). If the integrand in Eq. \eqref{eq:golden} dies
sufficiently fast as a function of \((t_1-t_2)\), then the decay rate
of the system out of the state \(\ket{2}\) can be defined as
\begin{equation}
  \begin{split}
    \tau^{-1}&=\frac{1}{4}\frac{\Delta_0^2}{\Delta_t^2}\int_{-\infty}^{\infty}
    \sum_k\lambda_k^2e^{i(\Delta_t-\omega_k)t}n(\omega_k)\dd{t}\\
    &=\frac{1}{4}\frac{\Delta_0^2}{\Delta_t^2}\sum_k2\pi\lambda_k^2
    \delta(\Delta_t-\omega_k)n(\omega_k)\\
    &=\pi\alpha\frac{\Delta_0^2}{\Delta_t}\exp(-\Delta_t/\omega_c)n(\Delta_t),
  \end{split}
\label{eq:boson-absorption}
\end{equation}
where \(n(\omega)=(e^{\beta\omega}-1)^{-1}\) is the Bose-Einstein
distribution function.

Similarly, if only a single phonon emission is considered in the emission
process then in \(\bar{H}_{SB}\) only the term
\(\frac{1}{2}\frac{\Delta_0}{\Delta_t}\sum_k\sigma_-b_k^{\dag}\).
Therefore the decay rate of the system out of the state \(\ket{1}\) is
\begin{equation}
  \tau^{-1}=\pi\alpha\frac{\Delta_0^2}{\Delta_t}
  \exp(-\Delta_t/\omega_c)[1+n(\Delta_t)].
  \label{eq:boson-emission}
\end{equation}
From Eq. \eqref{eq:boson-absorption} and \eqref{eq:boson-emission} it
can be seen that at zero temperature the decay rate out of the ground
state is zero, while the decay rate out of the excited state remains
finite, as they should be.

\subsection{Spin-Fermion Model}
In the spin-fermion model, for a fixed time \(t\), denoting \(vt\) in
Eq. \eqref{eq:lz-hamiltonian} by \(\varepsilon\) yields
\begin{equation}
  H_S=\frac{\varepsilon}{2}\sigma_z+\frac{\Delta_0}{2}\sigma_x.
\end{equation}
For simplicity, we assume \(\varepsilon\) is positive. This \(H_S\)
can be diagonalized by the same rotation as in
Eq. \eqref{eq:rotation}. Let
\(\Delta_t=\sqrt{\varepsilon^2+\Delta_0^2}\); then after the rotation
the system-bath coupling becomes
\begin{equation}
  \bar{H}_{SB}=\frac{1}{\Delta_t}\sum_{\alpha\beta;kq}V_{\alpha\beta}
  (\varepsilon\sigma_z-\Delta_0\sigma_x)c^{\dag}_{\alpha k}c_{\beta q}.
\end{equation}

Similarly, we can write \(\sigma_x\) in terms of \(\sigma_+\) and
\(\sigma_-\) as \(\sigma_x=\sigma_++\sigma_-\). In the absorption
process, if only a single electron jump is considered then only jumps
from the left lead to an unoccupied state with lower energy in the
right lead are permitted. The energy difference of the electron before
and after jump should be \(\Delta_t\). Therefore in \(\bar{H}_{SB}\)
only the term
\(\frac{1}{\Delta_t}\sum_{kq}V_{RL}\sigma_+c_{Rq}^{\dag}c_{Lk}\) is
relevant, and the golden rule formula for the decay rate of the system
out of the ground state can be written as
\begin{equation}
  \tau^{-1}=2\pi\lambda^2\frac{\Delta_0^2}{\Delta_t^2}(\Delta\mu-\Delta_t)
\end{equation}
for \(\Delta_t\le\Delta\mu\). When \(\Delta_t>\Delta\mu\), the decay
rate becomes zero; thus we have \(\Delta_c=\Delta\mu\).

The situation is more complex in the emission process. Two kinds
jumping are allowed: an electron in the left lead jumps to an
unoccupied state with higher energy in the right lead (the term
\(\frac{1}{\Delta_t}\sum_{kq}V_{RL}c_{Rk}^{\dag}c_{Lq}\) in
\(\bar{H}_{SB}\) is involved), and an electron in the right lead jumps
to an unoccupied state in the left lead (the term
\(\frac{1}{\Delta_t}\sum_{kq}V_{LR}c_{Lk}^{\dag}c_{Rq}\) in
\(\bar{H}_{SB}\) is involved). The energy difference before and after
the jump should be \(\Delta_t\). After summing up all possible jumps,
the decay rate in the former jump can be identified as
\begin{equation}
  \tau^{-1}=2\pi\lambda^2\frac{\Delta_0^2}{\Delta_t^2}\Delta\mu
\end{equation}
for \(\Delta_0\le\Delta_t\le\frac{1}{2}(D-\Delta\mu)\) and

\begin{equation}
  \tau^{-1}=2\pi\lambda^2\frac{\Delta_0^2}{\Delta_t^2}
  [\frac{1}{2}(D+\Delta\mu)-\Delta_t]
\end{equation}
for
\(\frac{1}{2}(D-\Delta\mu)<\Delta_t\le\frac{1}{2}(D+\Delta\mu)\). The
decay rate in the latter jump is
\begin{equation}
  \tau^{-1}=2\pi\lambda^2\frac{\Delta_0^2}{\Delta_t^2}(\Delta_t-\Delta\mu)
\end{equation}
for \(\Delta\mu<\Delta_t\le D\). The decay rate becomes zero under
other conditions; therefore \(\Delta_c=D\).


\begin{thebibliography}{66}%
\makeatletter
\providecommand \@ifxundefined [1]{%
 \@ifx{#1\undefined}
}%
\providecommand \@ifnum [1]{%
 \ifnum #1\expandafter \@firstoftwo
 \else \expandafter \@secondoftwo
 \fi
}%
\providecommand \@ifx [1]{%
 \ifx #1\expandafter \@firstoftwo
 \else \expandafter \@secondoftwo
 \fi
}%
\providecommand \natexlab [1]{#1}%
\providecommand \enquote  [1]{``#1''}%
\providecommand \bibnamefont  [1]{#1}%
\providecommand \bibfnamefont [1]{#1}%
\providecommand \citenamefont [1]{#1}%
\providecommand \href@noop [0]{\@secondoftwo}%
\providecommand \href [0]{\begingroup \@sanitize@url \@href}%
\providecommand \@href[1]{\@@startlink{#1}\@@href}%
\providecommand \@@href[1]{\endgroup#1\@@endlink}%
\providecommand \@sanitize@url [0]{\catcode `\\12\catcode `\$12\catcode
  `\&12\catcode `\#12\catcode `\^12\catcode `\_12\catcode `\%12\relax}%
\providecommand \@@startlink[1]{}%
\providecommand \@@endlink[0]{}%
\providecommand \url  [0]{\begingroup\@sanitize@url \@url }%
\providecommand \@url [1]{\endgroup\@href {#1}{\urlprefix }}%
\providecommand \urlprefix  [0]{URL }%
\providecommand \Eprint [0]{\href }%
\providecommand \doibase [0]{http://dx.doi.org/}%
\providecommand \selectlanguage [0]{\@gobble}%
\providecommand \bibinfo  [0]{\@secondoftwo}%
\providecommand \bibfield  [0]{\@secondoftwo}%
\providecommand \translation [1]{[#1]}%
\providecommand \BibitemOpen [0]{}%
\providecommand \bibitemStop [0]{}%
\providecommand \bibitemNoStop [0]{.\EOS\space}%
\providecommand \EOS [0]{\spacefactor3000\relax}%
\providecommand \BibitemShut  [1]{\csname bibitem#1\endcsname}%
\let\auto@bib@innerbib\@empty
\bibitem [{\citenamefont {Leggett}\ \emph {et~al.}(1987)\citenamefont
  {Leggett}, \citenamefont {Chakravarty}, \citenamefont {Dorsey}, \citenamefont
  {Fisher}, \citenamefont {Garg},\ and\ \citenamefont
  {Zwerger}}]{leggett1987-dynamics}%
  \BibitemOpen
  \bibfield  {author} {\bibinfo {author} {\bibfnamefont {A.~J.}\ \bibnamefont
  {Leggett}}, \bibinfo {author} {\bibfnamefont {S.}~\bibnamefont
  {Chakravarty}}, \bibinfo {author} {\bibfnamefont {A.~T.}\ \bibnamefont
  {Dorsey}}, \bibinfo {author} {\bibfnamefont {M.~P.~A.}\ \bibnamefont
  {Fisher}}, \bibinfo {author} {\bibfnamefont {A.}~\bibnamefont {Garg}}, \ and\
  \bibinfo {author} {\bibfnamefont {W.}~\bibnamefont {Zwerger}},\ }\href
  {\doibase 10.1103/revmodphys.59.1} {\bibfield  {journal} {\bibinfo  {journal}
  {Reviews of Modern Physics}\ }\textbf {\bibinfo {volume} {59}},\ \bibinfo
  {pages} {1} (\bibinfo {year} {1987})}\BibitemShut {NoStop}%
\bibitem [{\citenamefont {Weiss}(1993)}]{weiss1993-quantum}%
  \BibitemOpen
  \bibfield  {author} {\bibinfo {author} {\bibfnamefont {U.}~\bibnamefont
  {Weiss}},\ }\href {\doibase 10.1142/8334} {\emph {\bibinfo {title} {Quantum
  Dissipative Systems}}}\ (\bibinfo  {publisher} {World Scientific},\ \bibinfo
  {address} {Singapore},\ \bibinfo {year} {1993})\BibitemShut {NoStop}%
\bibitem [{\citenamefont {Hund}(1927)}]{hund1927-on}%
  \BibitemOpen
  \bibfield  {author} {\bibinfo {author} {\bibfnamefont {F.}~\bibnamefont
  {Hund}},\ }\href@noop {} {\bibfield  {journal} {\bibinfo  {journal}
  {Zeitschrift fur Physik}\ }\textbf {\bibinfo {volume} {43}},\ \bibinfo
  {pages} {805} (\bibinfo {year} {1927})}\BibitemShut {NoStop}%
\bibitem [{\citenamefont {Oppenheimer}(1928)}]{oppenheimer1928-three}%
  \BibitemOpen
  \bibfield  {author} {\bibinfo {author} {\bibfnamefont {J.~R.}\ \bibnamefont
  {Oppenheimer}},\ }\href {\doibase 10.1103/physrev.31.66} {\bibfield
  {journal} {\bibinfo  {journal} {Physical Review}\ }\textbf {\bibinfo {volume}
  {31}},\ \bibinfo {pages} {66} (\bibinfo {year} {1928})}\BibitemShut {NoStop}%
\bibitem [{\citenamefont {Landau}(1932)}]{landau1932-on}%
  \BibitemOpen
  \bibfield  {author} {\bibinfo {author} {\bibfnamefont {L.~D.}\ \bibnamefont
  {Landau}},\ }\href
  {https://www.sciencedirect.com/book/9780080105864/collected-papers-of-ld-landau}
  {\bibfield  {journal} {\bibinfo  {journal} {Physikalische Zeitschrift der
  Sowjetunion}\ }\textbf {\bibinfo {volume} {2}},\ \bibinfo {pages} {46}
  (\bibinfo {year} {1932})}\BibitemShut {NoStop}%
\bibitem [{\citenamefont {Zener}(1932)}]{zener1932-non}%
  \BibitemOpen
  \bibfield  {author} {\bibinfo {author} {\bibfnamefont {C.}~\bibnamefont
  {Zener}},\ }\href {\doibase 10.1098/rspa.1932.0165} {\bibfield  {journal}
  {\bibinfo  {journal} {Proceedings of the Royal Society A}\ }\textbf {\bibinfo
  {volume} {137}},\ \bibinfo {pages} {696} (\bibinfo {year}
  {1932})}\BibitemShut {NoStop}%
\bibitem [{\citenamefont {St\"uckelberg}(1932)}]{stueckelberg1932-theorie}%
  \BibitemOpen
  \bibfield  {author} {\bibinfo {author} {\bibfnamefont {E.}~\bibnamefont
  {St\"uckelberg}},\ }\href {\doibase 10.5169/seals-110177} {\bibfield
  {journal} {\bibinfo  {journal} {Helvetica Physica Acta}\ }\textbf {\bibinfo
  {volume} {5}},\ \bibinfo {pages} {369} (\bibinfo {year} {1932})}\BibitemShut
  {NoStop}%
\bibitem [{\citenamefont {Majorana}(1932)}]{majorana1932-atomi}%
  \BibitemOpen
  \bibfield  {author} {\bibinfo {author} {\bibfnamefont {E.}~\bibnamefont
  {Majorana}},\ }\href {\doibase 10.1007/BF02960953} {\bibfield  {journal}
  {\bibinfo  {journal} {Nuovo Cimento}\ }\textbf {\bibinfo {volume} {9}},\
  \bibinfo {pages} {43} (\bibinfo {year} {1932})}\BibitemShut {NoStop}%
\bibitem [{\citenamefont {May}\ and\ \citenamefont
  {K\"uhn}(2011)}]{May2011-charge}%
  \BibitemOpen
  \bibfield  {author} {\bibinfo {author} {\bibfnamefont {V.}~\bibnamefont
  {May}}\ and\ \bibinfo {author} {\bibfnamefont {O.}~\bibnamefont {K\"uhn}},\
  }\href {\doibase 10.1002/9783527633791} {\emph {\bibinfo {title} {Charge and
  Energy Transfer Dynamics in Molecular Systems}}}\ (\bibinfo  {publisher}
  {Wiley-VCH verlag, Berlin},\ \bibinfo {year} {2011})\BibitemShut {NoStop}%
\bibitem [{\citenamefont {Thiel}(1990)}]{thiel1990-landau}%
  \BibitemOpen
  \bibfield  {author} {\bibinfo {author} {\bibfnamefont {A.}~\bibnamefont
  {Thiel}},\ }\href {\doibase 10.1088/0954-3899/16/7/004} {\bibfield  {journal}
  {\bibinfo  {journal} {Journal of Physics G: Nuclear and Particle Physics}\
  }\textbf {\bibinfo {volume} {16}},\ \bibinfo {pages} {867} (\bibinfo {year}
  {1990})}\BibitemShut {NoStop}%
\bibitem [{\citenamefont {Harmin}\ and\ \citenamefont
  {Price}(1994)}]{harmin1994-incoherent}%
  \BibitemOpen
  \bibfield  {author} {\bibinfo {author} {\bibfnamefont {D.~A.}\ \bibnamefont
  {Harmin}}\ and\ \bibinfo {author} {\bibfnamefont {P.~N.}\ \bibnamefont
  {Price}},\ }\href {\doibase 10.1103/physreva.49.1933} {\bibfield  {journal}
  {\bibinfo  {journal} {Physical Review A}\ }\textbf {\bibinfo {volume} {49}},\
  \bibinfo {pages} {1933} (\bibinfo {year} {1994})}\BibitemShut {NoStop}%
\bibitem [{\citenamefont {Xie}\ and\ \citenamefont
  {Domcke}(2017)}]{xie2017-accuracy}%
  \BibitemOpen
  \bibfield  {author} {\bibinfo {author} {\bibfnamefont {W.}~\bibnamefont
  {Xie}}\ and\ \bibinfo {author} {\bibfnamefont {W.}~\bibnamefont {Domcke}},\
  }\href {\doibase 10.1063/1.5006788} {\bibfield  {journal} {\bibinfo
  {journal} {The Journal of Chemical Physics}\ }\textbf {\bibinfo {volume}
  {147}},\ \bibinfo {pages} {184114} (\bibinfo {year} {2017})}\BibitemShut
  {NoStop}%
\bibitem [{\citenamefont {Sillanp\"a\"a}\ \emph {et~al.}(2006)\citenamefont
  {Sillanp\"a\"a}, \citenamefont {Lehtinen}, \citenamefont {Paila},
  \citenamefont {Makhlin},\ and\ \citenamefont
  {Hakonen}}]{sillanpaa2006-continuous}%
  \BibitemOpen
  \bibfield  {author} {\bibinfo {author} {\bibfnamefont {M.}~\bibnamefont
  {Sillanp\"a\"a}}, \bibinfo {author} {\bibfnamefont {T.}~\bibnamefont
  {Lehtinen}}, \bibinfo {author} {\bibfnamefont {A.}~\bibnamefont {Paila}},
  \bibinfo {author} {\bibfnamefont {Y.}~\bibnamefont {Makhlin}}, \ and\
  \bibinfo {author} {\bibfnamefont {P.}~\bibnamefont {Hakonen}},\ }\href
  {\doibase 10.1103/physrevlett.96.187002} {\bibfield  {journal} {\bibinfo
  {journal} {Physical Review Letters}\ }\textbf {\bibinfo {volume} {96}},\
  \bibinfo {pages} {187002} (\bibinfo {year} {2006})}\BibitemShut {NoStop}%
\bibitem [{\citenamefont {Berns}\ \emph {et~al.}(2008)\citenamefont {Berns},
  \citenamefont {Rudner}, \citenamefont {Valenzuela}, \citenamefont {Berggren},
  \citenamefont {Oliver}, \citenamefont {Levitov},\ and\ \citenamefont
  {Orlando}}]{berns2008-amplitude}%
  \BibitemOpen
  \bibfield  {author} {\bibinfo {author} {\bibfnamefont {D.~M.}\ \bibnamefont
  {Berns}}, \bibinfo {author} {\bibfnamefont {M.~S.}\ \bibnamefont {Rudner}},
  \bibinfo {author} {\bibfnamefont {S.~O.}\ \bibnamefont {Valenzuela}},
  \bibinfo {author} {\bibfnamefont {K.~K.}\ \bibnamefont {Berggren}}, \bibinfo
  {author} {\bibfnamefont {W.~D.}\ \bibnamefont {Oliver}}, \bibinfo {author}
  {\bibfnamefont {L.~S.}\ \bibnamefont {Levitov}}, \ and\ \bibinfo {author}
  {\bibfnamefont {T.~P.}\ \bibnamefont {Orlando}},\ }\href {\doibase
  10.1038/nature07262} {\bibfield  {journal} {\bibinfo  {journal} {Nature}\
  }\textbf {\bibinfo {volume} {455}},\ \bibinfo {pages} {51} (\bibinfo {year}
  {2008})}\BibitemShut {NoStop}%
\bibitem [{\citenamefont {DeRaedt}\ \emph {et~al.}(1997)\citenamefont
  {DeRaedt}, \citenamefont {Miyashita}, \citenamefont {Saito}, \citenamefont
  {Garc{\'i}a-Pablos},\ and\ \citenamefont {Garc{\'i}a}}]{raedt1997-theory}%
  \BibitemOpen
  \bibfield  {author} {\bibinfo {author} {\bibfnamefont {H.}~\bibnamefont
  {DeRaedt}}, \bibinfo {author} {\bibfnamefont {S.}~\bibnamefont {Miyashita}},
  \bibinfo {author} {\bibfnamefont {K.}~\bibnamefont {Saito}}, \bibinfo
  {author} {\bibfnamefont {D.}~\bibnamefont {Garc{\'i}a-Pablos}}, \ and\
  \bibinfo {author} {\bibfnamefont {N.}~\bibnamefont {Garc{\'i}a}},\ }\href
  {\doibase 10.1103/physrevb.56.11761} {\bibfield  {journal} {\bibinfo
  {journal} {Physical Review B}\ }\textbf {\bibinfo {volume} {56}},\ \bibinfo
  {pages} {11761} (\bibinfo {year} {1997})}\BibitemShut {NoStop}%
\bibitem [{\citenamefont {Wernsdorfer}(1999)}]{wernsdorfer1999-quantum}%
  \BibitemOpen
  \bibfield  {author} {\bibinfo {author} {\bibfnamefont {W.}~\bibnamefont
  {Wernsdorfer}},\ }\href {\doibase 10.1126/science.284.5411.133} {\bibfield
  {journal} {\bibinfo  {journal} {Science}\ }\textbf {\bibinfo {volume}
  {284}},\ \bibinfo {pages} {133} (\bibinfo {year} {1999})}\BibitemShut
  {NoStop}%
\bibitem [{\citenamefont {Spreeuw}\ \emph {et~al.}(1990)\citenamefont
  {Spreeuw}, \citenamefont {van Druten}, \citenamefont {Beijersbergen},
  \citenamefont {Eliel},\ and\ \citenamefont
  {Woerdman}}]{spreeuw1990-classical}%
  \BibitemOpen
  \bibfield  {author} {\bibinfo {author} {\bibfnamefont {R.~J.~C.}\
  \bibnamefont {Spreeuw}}, \bibinfo {author} {\bibfnamefont {N.~J.}\
  \bibnamefont {van Druten}}, \bibinfo {author} {\bibfnamefont {M.~W.}\
  \bibnamefont {Beijersbergen}}, \bibinfo {author} {\bibfnamefont {E.~R.}\
  \bibnamefont {Eliel}}, \ and\ \bibinfo {author} {\bibfnamefont {J.~P.}\
  \bibnamefont {Woerdman}},\ }\href {\doibase 10.1103/physrevlett.65.2642}
  {\bibfield  {journal} {\bibinfo  {journal} {Physical Review Letters}\
  }\textbf {\bibinfo {volume} {65}},\ \bibinfo {pages} {2642} (\bibinfo {year}
  {1990})}\BibitemShut {NoStop}%
\bibitem [{\citenamefont {Bouwmeester}\ \emph {et~al.}(1995)\citenamefont
  {Bouwmeester}, \citenamefont {Dekker}, \citenamefont {v.~Dorsselaer},
  \citenamefont {Schrama}, \citenamefont {Visser},\ and\ \citenamefont
  {Woerdman}}]{bouwmeester1995-observation}%
  \BibitemOpen
  \bibfield  {author} {\bibinfo {author} {\bibfnamefont {D.}~\bibnamefont
  {Bouwmeester}}, \bibinfo {author} {\bibfnamefont {N.~H.}\ \bibnamefont
  {Dekker}}, \bibinfo {author} {\bibfnamefont {F.~E.}\ \bibnamefont
  {v.~Dorsselaer}}, \bibinfo {author} {\bibfnamefont {C.~A.}\ \bibnamefont
  {Schrama}}, \bibinfo {author} {\bibfnamefont {P.~M.}\ \bibnamefont {Visser}},
  \ and\ \bibinfo {author} {\bibfnamefont {J.~P.}\ \bibnamefont {Woerdman}},\
  }\href {\doibase 10.1103/physreva.51.646} {\bibfield  {journal} {\bibinfo
  {journal} {Physical Review A}\ }\textbf {\bibinfo {volume} {51}},\ \bibinfo
  {pages} {646} (\bibinfo {year} {1995})}\BibitemShut {NoStop}%
\bibitem [{\citenamefont {Witthaut}\ \emph {et~al.}(2006)\citenamefont
  {Witthaut}, \citenamefont {Graefe},\ and\ \citenamefont
  {Korsch}}]{witthaut2006-towards}%
  \BibitemOpen
  \bibfield  {author} {\bibinfo {author} {\bibfnamefont {D.}~\bibnamefont
  {Witthaut}}, \bibinfo {author} {\bibfnamefont {E.~M.}\ \bibnamefont
  {Graefe}}, \ and\ \bibinfo {author} {\bibfnamefont {H.~J.}\ \bibnamefont
  {Korsch}},\ }\href {\doibase 10.1103/physreva.73.063609} {\bibfield
  {journal} {\bibinfo  {journal} {Physical Review A}\ }\textbf {\bibinfo
  {volume} {73}},\ \bibinfo {pages} {063609} (\bibinfo {year}
  {2006})}\BibitemShut {NoStop}%
\bibitem [{\citenamefont {Zenesini}\ \emph {et~al.}(2009)\citenamefont
  {Zenesini}, \citenamefont {Lignier}, \citenamefont {Tayebirad}, \citenamefont
  {Radogostowicz}, \citenamefont {Ciampini}, \citenamefont {Mannella},
  \citenamefont {Wimberger}, \citenamefont {Morsch},\ and\ \citenamefont
  {Arimondo}}]{zenesini2009-time}%
  \BibitemOpen
  \bibfield  {author} {\bibinfo {author} {\bibfnamefont {A.}~\bibnamefont
  {Zenesini}}, \bibinfo {author} {\bibfnamefont {H.}~\bibnamefont {Lignier}},
  \bibinfo {author} {\bibfnamefont {G.}~\bibnamefont {Tayebirad}}, \bibinfo
  {author} {\bibfnamefont {J.}~\bibnamefont {Radogostowicz}}, \bibinfo {author}
  {\bibfnamefont {D.}~\bibnamefont {Ciampini}}, \bibinfo {author}
  {\bibfnamefont {R.}~\bibnamefont {Mannella}}, \bibinfo {author}
  {\bibfnamefont {S.}~\bibnamefont {Wimberger}}, \bibinfo {author}
  {\bibfnamefont {O.}~\bibnamefont {Morsch}}, \ and\ \bibinfo {author}
  {\bibfnamefont {E.}~\bibnamefont {Arimondo}},\ }\href {\doibase
  10.1103/physrevlett.103.090403} {\bibfield  {journal} {\bibinfo  {journal}
  {Physical Review Letters}\ }\textbf {\bibinfo {volume} {103}},\ \bibinfo
  {pages} {090403} (\bibinfo {year} {2009})}\BibitemShut {NoStop}%
\bibitem [{\citenamefont {Olson}\ \emph {et~al.}(2014)\citenamefont {Olson},
  \citenamefont {Wang}, \citenamefont {Niffenegger}, \citenamefont {Li},
  \citenamefont {Greene},\ and\ \citenamefont {Chen}}]{olson2014-tunable}%
  \BibitemOpen
  \bibfield  {author} {\bibinfo {author} {\bibfnamefont {A.~J.}\ \bibnamefont
  {Olson}}, \bibinfo {author} {\bibfnamefont {S.-J.}\ \bibnamefont {Wang}},
  \bibinfo {author} {\bibfnamefont {R.~J.}\ \bibnamefont {Niffenegger}},
  \bibinfo {author} {\bibfnamefont {C.-H.}\ \bibnamefont {Li}}, \bibinfo
  {author} {\bibfnamefont {C.~H.}\ \bibnamefont {Greene}}, \ and\ \bibinfo
  {author} {\bibfnamefont {Y.~P.}\ \bibnamefont {Chen}},\ }\href {\doibase
  10.1103/physreva.90.013616} {\bibfield  {journal} {\bibinfo  {journal}
  {Physical Review A}\ }\textbf {\bibinfo {volume} {90}},\ \bibinfo {pages}
  {013616} (\bibinfo {year} {2014})}\BibitemShut {NoStop}%
\bibitem [{\citenamefont {Ankerhold}\ and\ \citenamefont
  {Grabert}(2003)}]{ankerhold2003-enhancement}%
  \BibitemOpen
  \bibfield  {author} {\bibinfo {author} {\bibfnamefont {J.}~\bibnamefont
  {Ankerhold}}\ and\ \bibinfo {author} {\bibfnamefont {H.}~\bibnamefont
  {Grabert}},\ }\href {\doibase 10.1103/physrevlett.91.016803} {\bibfield
  {journal} {\bibinfo  {journal} {Physical Review Letters}\ }\textbf {\bibinfo
  {volume} {91}},\ \bibinfo {pages} {016803} (\bibinfo {year}
  {2003})}\BibitemShut {NoStop}%
\bibitem [{\citenamefont {Izmalkov}\ \emph {et~al.}(2004)\citenamefont
  {Izmalkov}, \citenamefont {Grajcar}, \citenamefont {Il'ichev}, \citenamefont
  {Oukhanski}, \citenamefont {Wagner}, \citenamefont {Meyer}, \citenamefont
  {Krech}, \citenamefont {Amin}, \citenamefont {van~den Brink},\ and\
  \citenamefont {Zagoskin}}]{izmalkov2004-observation}%
  \BibitemOpen
  \bibfield  {author} {\bibinfo {author} {\bibfnamefont {A.}~\bibnamefont
  {Izmalkov}}, \bibinfo {author} {\bibfnamefont {M.}~\bibnamefont {Grajcar}},
  \bibinfo {author} {\bibfnamefont {E.}~\bibnamefont {Il'ichev}}, \bibinfo
  {author} {\bibfnamefont {N.}~\bibnamefont {Oukhanski}}, \bibinfo {author}
  {\bibfnamefont {T.}~\bibnamefont {Wagner}}, \bibinfo {author} {\bibfnamefont
  {H.-G.}\ \bibnamefont {Meyer}}, \bibinfo {author} {\bibfnamefont
  {W.}~\bibnamefont {Krech}}, \bibinfo {author} {\bibfnamefont {M.~H.~S.}\
  \bibnamefont {Amin}}, \bibinfo {author} {\bibfnamefont {A.~M.}\ \bibnamefont
  {van~den Brink}}, \ and\ \bibinfo {author} {\bibfnamefont {A.~M.}\
  \bibnamefont {Zagoskin}},\ }\href {\doibase 10.1209/epl/i2003-10200-6}
  {\bibfield  {journal} {\bibinfo  {journal} {Europhysics Letters}\ }\textbf
  {\bibinfo {volume} {65}},\ \bibinfo {pages} {844} (\bibinfo {year}
  {2004})}\BibitemShut {NoStop}%
\bibitem [{\citenamefont {Wubs}\ \emph {et~al.}(2005)\citenamefont {Wubs},
  \citenamefont {Saito}, \citenamefont {Kohler}, \citenamefont {Kayanuma},\
  and\ \citenamefont {H{\"a}nggi}}]{wubs2005-landau}%
  \BibitemOpen
  \bibfield  {author} {\bibinfo {author} {\bibfnamefont {M.}~\bibnamefont
  {Wubs}}, \bibinfo {author} {\bibfnamefont {K.}~\bibnamefont {Saito}},
  \bibinfo {author} {\bibfnamefont {S.}~\bibnamefont {Kohler}}, \bibinfo
  {author} {\bibfnamefont {Y.}~\bibnamefont {Kayanuma}}, \ and\ \bibinfo
  {author} {\bibfnamefont {P.}~\bibnamefont {H{\"a}nggi}},\ }\href {\doibase
  10.1088/1367-2630/7/1/218} {\bibfield  {journal} {\bibinfo  {journal} {New
  Journal of Physics}\ }\textbf {\bibinfo {volume} {7}},\ \bibinfo {pages}
  {218} (\bibinfo {year} {2005})}\BibitemShut {NoStop}%
\bibitem [{\citenamefont {Oliver}\ \emph {et~al.}(2005)\citenamefont {Oliver},
  \citenamefont {Yu}, \citenamefont {Lee}, \citenamefont {Berggren},
  \citenamefont {Levitov},\ and\ \citenamefont {Orlando}}]{oliver2005-mach}%
  \BibitemOpen
  \bibfield  {author} {\bibinfo {author} {\bibfnamefont {W.~D.}\ \bibnamefont
  {Oliver}}, \bibinfo {author} {\bibfnamefont {Y.}~\bibnamefont {Yu}}, \bibinfo
  {author} {\bibfnamefont {J.~C.}\ \bibnamefont {Lee}}, \bibinfo {author}
  {\bibfnamefont {K.~K.}\ \bibnamefont {Berggren}}, \bibinfo {author}
  {\bibfnamefont {L.~S.}\ \bibnamefont {Levitov}}, \ and\ \bibinfo {author}
  {\bibfnamefont {T.~P.}\ \bibnamefont {Orlando}},\ }\href {\doibase
  10.1126/science.1119678} {\bibfield  {journal} {\bibinfo  {journal}
  {Science}\ }\textbf {\bibinfo {volume} {310}},\ \bibinfo {pages} {1653}
  (\bibinfo {year} {2005})}\BibitemShut {NoStop}%
\bibitem [{\citenamefont {Wei}\ \emph {et~al.}(2008)\citenamefont {Wei},
  \citenamefont {Johansson}, \citenamefont {Cen}, \citenamefont {Ashhab},\ and\
  \citenamefont {Nori}}]{wei2008-controllable}%
  \BibitemOpen
  \bibfield  {author} {\bibinfo {author} {\bibfnamefont {L.~F.}\ \bibnamefont
  {Wei}}, \bibinfo {author} {\bibfnamefont {J.~R.}\ \bibnamefont {Johansson}},
  \bibinfo {author} {\bibfnamefont {L.~X.}\ \bibnamefont {Cen}}, \bibinfo
  {author} {\bibfnamefont {S.}~\bibnamefont {Ashhab}}, \ and\ \bibinfo {author}
  {\bibfnamefont {F.}~\bibnamefont {Nori}},\ }\href {\doibase
  10.1103/physrevlett.100.113601} {\bibfield  {journal} {\bibinfo  {journal}
  {Physical Review Letters}\ }\textbf {\bibinfo {volume} {100}},\ \bibinfo
  {pages} {113601} (\bibinfo {year} {2008})}\BibitemShut {NoStop}%
\bibitem [{\citenamefont {Fuchs}\ \emph {et~al.}(2011)\citenamefont {Fuchs},
  \citenamefont {Burkard}, \citenamefont {Klimov},\ and\ \citenamefont
  {Awschalom}}]{fuchs2011-quantum}%
  \BibitemOpen
  \bibfield  {author} {\bibinfo {author} {\bibfnamefont {G.~D.}\ \bibnamefont
  {Fuchs}}, \bibinfo {author} {\bibfnamefont {G.}~\bibnamefont {Burkard}},
  \bibinfo {author} {\bibfnamefont {P.~V.}\ \bibnamefont {Klimov}}, \ and\
  \bibinfo {author} {\bibfnamefont {D.~D.}\ \bibnamefont {Awschalom}},\ }\href
  {\doibase 10.1038/nphys2026} {\bibfield  {journal} {\bibinfo  {journal}
  {Nature Physics}\ }\textbf {\bibinfo {volume} {7}},\ \bibinfo {pages} {789}
  (\bibinfo {year} {2011})}\BibitemShut {NoStop}%
\bibitem [{\citenamefont {Shytov}\ \emph {et~al.}(2003)\citenamefont {Shytov},
  \citenamefont {Ivanov},\ and\ \citenamefont
  {Feigel'man}}]{shytov2003-landau}%
  \BibitemOpen
  \bibfield  {author} {\bibinfo {author} {\bibfnamefont {A.~V.}\ \bibnamefont
  {Shytov}}, \bibinfo {author} {\bibfnamefont {D.~A.}\ \bibnamefont {Ivanov}},
  \ and\ \bibinfo {author} {\bibfnamefont {M.~V.}\ \bibnamefont {Feigel'man}},\
  }\href {\doibase 10.1140/epjb/e2003-00343-8} {\bibfield  {journal} {\bibinfo
  {journal} {The European Physical Journal B}\ }\textbf {\bibinfo {volume}
  {36}},\ \bibinfo {pages} {263} (\bibinfo {year} {2003})}\BibitemShut
  {NoStop}%
\bibitem [{\citenamefont {Izmalkov}\ \emph {et~al.}(2008)\citenamefont
  {Izmalkov}, \citenamefont {van~der Ploeg}, \citenamefont {Shevchenko},
  \citenamefont {Grajcar}, \citenamefont {Il'ichev}, \citenamefont
  {H{\"u}bner}, \citenamefont {Omelyanchouk},\ and\ \citenamefont
  {Meyer}}]{izmalkov2008-consistency}%
  \BibitemOpen
  \bibfield  {author} {\bibinfo {author} {\bibfnamefont {A.}~\bibnamefont
  {Izmalkov}}, \bibinfo {author} {\bibfnamefont {S.~H.~W.}\ \bibnamefont
  {van~der Ploeg}}, \bibinfo {author} {\bibfnamefont {S.~N.}\ \bibnamefont
  {Shevchenko}}, \bibinfo {author} {\bibfnamefont {M.}~\bibnamefont {Grajcar}},
  \bibinfo {author} {\bibfnamefont {E.}~\bibnamefont {Il'ichev}}, \bibinfo
  {author} {\bibfnamefont {U.}~\bibnamefont {H{\"u}bner}}, \bibinfo {author}
  {\bibfnamefont {A.~N.}\ \bibnamefont {Omelyanchouk}}, \ and\ \bibinfo
  {author} {\bibfnamefont {H.-G.}\ \bibnamefont {Meyer}},\ }\href {\doibase
  10.1103/physrevlett.101.017003} {\bibfield  {journal} {\bibinfo  {journal}
  {Physical Review Letters}\ }\textbf {\bibinfo {volume} {101}},\ \bibinfo
  {pages} {017003} (\bibinfo {year} {2008})}\BibitemShut {NoStop}%
\bibitem [{\citenamefont {Shevchenko}\ \emph {et~al.}(2010)\citenamefont
  {Shevchenko}, \citenamefont {Ashhab},\ and\ \citenamefont
  {Nori}}]{shevchenko2010-landau}%
  \BibitemOpen
  \bibfield  {author} {\bibinfo {author} {\bibfnamefont {S.}~\bibnamefont
  {Shevchenko}}, \bibinfo {author} {\bibfnamefont {S.}~\bibnamefont {Ashhab}},
  \ and\ \bibinfo {author} {\bibfnamefont {F.}~\bibnamefont {Nori}},\ }\href
  {\doibase 10.1016/j.physrep.2010.03.002} {\bibfield  {journal} {\bibinfo
  {journal} {Physics Reports}\ }\textbf {\bibinfo {volume} {492}},\ \bibinfo
  {pages} {1} (\bibinfo {year} {2010})}\BibitemShut {NoStop}%
\bibitem [{\citenamefont {Neilinger}\ \emph {et~al.}(2016)\citenamefont
  {Neilinger}, \citenamefont {Shevchenko}, \citenamefont {Bog{\'a}r},
  \citenamefont {Reh{\'a}k}, \citenamefont {Oelsner}, \citenamefont {Karpov},
  \citenamefont {H{\"u}bner}, \citenamefont {Astafiev}, \citenamefont
  {Grajcar},\ and\ \citenamefont {Il'ichev}}]{neilinger2016-landau}%
  \BibitemOpen
  \bibfield  {author} {\bibinfo {author} {\bibfnamefont {P.}~\bibnamefont
  {Neilinger}}, \bibinfo {author} {\bibfnamefont {S.~N.}\ \bibnamefont
  {Shevchenko}}, \bibinfo {author} {\bibfnamefont {J.}~\bibnamefont
  {Bog{\'a}r}}, \bibinfo {author} {\bibfnamefont {M.}~\bibnamefont
  {Reh{\'a}k}}, \bibinfo {author} {\bibfnamefont {G.}~\bibnamefont {Oelsner}},
  \bibinfo {author} {\bibfnamefont {D.~S.}\ \bibnamefont {Karpov}}, \bibinfo
  {author} {\bibfnamefont {U.}~\bibnamefont {H{\"u}bner}}, \bibinfo {author}
  {\bibfnamefont {O.}~\bibnamefont {Astafiev}}, \bibinfo {author}
  {\bibfnamefont {M.}~\bibnamefont {Grajcar}}, \ and\ \bibinfo {author}
  {\bibfnamefont {E.}~\bibnamefont {Il'ichev}},\ }\href {\doibase
  10.1103/physrevb.94.094519} {\bibfield  {journal} {\bibinfo  {journal}
  {Physical Review B}\ }\textbf {\bibinfo {volume} {94}},\ \bibinfo {pages}
  {094519} (\bibinfo {year} {2016})}\BibitemShut {NoStop}%
\bibitem [{\citenamefont {Chatterjee}\ \emph {et~al.}(2018)\citenamefont
  {Chatterjee}, \citenamefont {Shevchenko}, \citenamefont {Barraud},
  \citenamefont {Otxoa}, \citenamefont {Nori}, \citenamefont {Morton},\ and\
  \citenamefont {Gonzalez-Zalba}}]{chatterjee2018-silicon}%
  \BibitemOpen
  \bibfield  {author} {\bibinfo {author} {\bibfnamefont {A.}~\bibnamefont
  {Chatterjee}}, \bibinfo {author} {\bibfnamefont {S.~N.}\ \bibnamefont
  {Shevchenko}}, \bibinfo {author} {\bibfnamefont {S.}~\bibnamefont {Barraud}},
  \bibinfo {author} {\bibfnamefont {R.~M.}\ \bibnamefont {Otxoa}}, \bibinfo
  {author} {\bibfnamefont {F.}~\bibnamefont {Nori}}, \bibinfo {author}
  {\bibfnamefont {J.~J.~L.}\ \bibnamefont {Morton}}, \ and\ \bibinfo {author}
  {\bibfnamefont {M.~F.}\ \bibnamefont {Gonzalez-Zalba}},\ }\href {\doibase
  10.1103/physrevb.97.045405} {\bibfield  {journal} {\bibinfo  {journal}
  {Physical Review B}\ }\textbf {\bibinfo {volume} {97}},\ \bibinfo {pages}
  {045405} (\bibinfo {year} {2018})}\BibitemShut {NoStop}%
\bibitem [{\citenamefont {Wang}\ \emph {et~al.}(2018)\citenamefont {Wang},
  \citenamefont {Huang}, \citenamefont {Liang},\ and\ \citenamefont
  {Hu}}]{wang2018-landau}%
  \BibitemOpen
  \bibfield  {author} {\bibinfo {author} {\bibfnamefont {Z.}~\bibnamefont
  {Wang}}, \bibinfo {author} {\bibfnamefont {W.~C.}\ \bibnamefont {Huang}},
  \bibinfo {author} {\bibfnamefont {Q.~F.}\ \bibnamefont {Liang}}, \ and\
  \bibinfo {author} {\bibfnamefont {X.}~\bibnamefont {Hu}},\ }\href {\doibase
  10.1038/s41598-018-26324-5} {\bibfield  {journal} {\bibinfo  {journal}
  {Scientific Reports}\ }\textbf {\bibinfo {volume} {8}},\ \bibinfo {pages}
  {7920} (\bibinfo {year} {2018})}\BibitemShut {NoStop}%
\bibitem [{\citenamefont {Kayanuma}(1984)}]{kayanuma1984-nonadiabatic}%
  \BibitemOpen
  \bibfield  {author} {\bibinfo {author} {\bibfnamefont {Y.}~\bibnamefont
  {Kayanuma}},\ }\href {\doibase 10.1143/jpsj.53.108} {\bibfield  {journal}
  {\bibinfo  {journal} {Journal of the Physical Society of Japan}\ }\textbf
  {\bibinfo {volume} {53}},\ \bibinfo {pages} {108} (\bibinfo {year}
  {1984})}\BibitemShut {NoStop}%
\bibitem [{\citenamefont {Grifoni}\ and\ \citenamefont
  {H\"anggi}(1998)}]{grifoni1998-driven}%
  \BibitemOpen
  \bibfield  {author} {\bibinfo {author} {\bibfnamefont {M.}~\bibnamefont
  {Grifoni}}\ and\ \bibinfo {author} {\bibfnamefont {P.}~\bibnamefont
  {H\"anggi}},\ }\href {\doibase 10.1016/S0370-1573(98)00022-2} {\bibfield
  {journal} {\bibinfo  {journal} {Physics Reports}\ }\textbf {\bibinfo {volume}
  {304}},\ \bibinfo {pages} {229} (\bibinfo {year} {1998})}\BibitemShut
  {NoStop}%
\bibitem [{\citenamefont {Wittig}(2005)}]{wittig2005-landau}%
  \BibitemOpen
  \bibfield  {author} {\bibinfo {author} {\bibfnamefont {C.}~\bibnamefont
  {Wittig}},\ }\href {\doibase 10.1021/jp040627u} {\bibfield  {journal}
  {\bibinfo  {journal} {The Journal of Physical Chemistry B}\ }\textbf
  {\bibinfo {volume} {109}},\ \bibinfo {pages} {8428} (\bibinfo {year}
  {2005})}\BibitemShut {NoStop}%
\bibitem [{\citenamefont {Gefen}\ \emph {et~al.}(1987)\citenamefont {Gefen},
  \citenamefont {Ben-Jacob},\ and\ \citenamefont {Caldeira}}]{gefen1987-zener}%
  \BibitemOpen
  \bibfield  {author} {\bibinfo {author} {\bibfnamefont {Y.}~\bibnamefont
  {Gefen}}, \bibinfo {author} {\bibfnamefont {E.}~\bibnamefont {Ben-Jacob}}, \
  and\ \bibinfo {author} {\bibfnamefont {A.~O.}\ \bibnamefont {Caldeira}},\
  }\href {\doibase 10.1103/physrevb.36.2770} {\bibfield  {journal} {\bibinfo
  {journal} {Physical Review B}\ }\textbf {\bibinfo {volume} {36}},\ \bibinfo
  {pages} {2770} (\bibinfo {year} {1987})}\BibitemShut {NoStop}%
\bibitem [{\citenamefont {Ao}\ and\ \citenamefont
  {Rammer}(1989)}]{ao1989-influence}%
  \BibitemOpen
  \bibfield  {author} {\bibinfo {author} {\bibfnamefont {P.}~\bibnamefont
  {Ao}}\ and\ \bibinfo {author} {\bibfnamefont {J.}~\bibnamefont {Rammer}},\
  }\href {\doibase 10.1103/physrevlett.62.3004} {\bibfield  {journal} {\bibinfo
   {journal} {Physical Review Letters}\ }\textbf {\bibinfo {volume} {62}},\
  \bibinfo {pages} {3004} (\bibinfo {year} {1989})}\BibitemShut {NoStop}%
\bibitem [{\citenamefont {Ao}\ and\ \citenamefont
  {Rammer}(1991)}]{ao1991-quantum}%
  \BibitemOpen
  \bibfield  {author} {\bibinfo {author} {\bibfnamefont {P.}~\bibnamefont
  {Ao}}\ and\ \bibinfo {author} {\bibfnamefont {J.}~\bibnamefont {Rammer}},\
  }\href {\doibase 10.1103/physrevb.43.5397} {\bibfield  {journal} {\bibinfo
  {journal} {Physical Review B}\ }\textbf {\bibinfo {volume} {43}},\ \bibinfo
  {pages} {5397} (\bibinfo {year} {1991})}\BibitemShut {NoStop}%
\bibitem [{\citenamefont {Kayanuma}\ and\ \citenamefont
  {Nakayama}(1998)}]{kayanuma1998-nonadiabatic}%
  \BibitemOpen
  \bibfield  {author} {\bibinfo {author} {\bibfnamefont {Y.}~\bibnamefont
  {Kayanuma}}\ and\ \bibinfo {author} {\bibfnamefont {H.}~\bibnamefont
  {Nakayama}},\ }\href {\doibase 10.1103/physrevb.57.13099} {\bibfield
  {journal} {\bibinfo  {journal} {Physical Review B}\ }\textbf {\bibinfo
  {volume} {57}},\ \bibinfo {pages} {13099} (\bibinfo {year}
  {1998})}\BibitemShut {NoStop}%
\bibitem [{\citenamefont {Kobayashi}\ \emph {et~al.}(1999)\citenamefont
  {Kobayashi}, \citenamefont {Hatano},\ and\ \citenamefont
  {Miyashita}}]{kobayashi1999-non}%
  \BibitemOpen
  \bibfield  {author} {\bibinfo {author} {\bibfnamefont {H.}~\bibnamefont
  {Kobayashi}}, \bibinfo {author} {\bibfnamefont {N.}~\bibnamefont {Hatano}}, \
  and\ \bibinfo {author} {\bibfnamefont {S.}~\bibnamefont {Miyashita}},\ }\href
  {\doibase 10.1016/s0378-4371(98)00475-0} {\bibfield  {journal} {\bibinfo
  {journal} {Physica A: Statistical Mechanics and its Applications}\ }\textbf
  {\bibinfo {volume} {265}},\ \bibinfo {pages} {565} (\bibinfo {year}
  {1999})}\BibitemShut {NoStop}%
\bibitem [{\citenamefont {Wubs}\ \emph {et~al.}(2006)\citenamefont {Wubs},
  \citenamefont {Saito}, \citenamefont {Kohler}, \citenamefont {H{\"a}nggi},\
  and\ \citenamefont {Kayanuma}}]{wubs2006-gauging}%
  \BibitemOpen
  \bibfield  {author} {\bibinfo {author} {\bibfnamefont {M.}~\bibnamefont
  {Wubs}}, \bibinfo {author} {\bibfnamefont {K.}~\bibnamefont {Saito}},
  \bibinfo {author} {\bibfnamefont {S.}~\bibnamefont {Kohler}}, \bibinfo
  {author} {\bibfnamefont {P.}~\bibnamefont {H{\"a}nggi}}, \ and\ \bibinfo
  {author} {\bibfnamefont {Y.}~\bibnamefont {Kayanuma}},\ }\href {\doibase
  10.1103/physrevlett.97.200404} {\bibfield  {journal} {\bibinfo  {journal}
  {Physical Review Letters}\ }\textbf {\bibinfo {volume} {97}},\ \bibinfo
  {pages} {200404} (\bibinfo {year} {2006})}\BibitemShut {NoStop}%
\bibitem [{\citenamefont {Saito}\ \emph {et~al.}(2007)\citenamefont {Saito},
  \citenamefont {Wubs}, \citenamefont {Kohler}, \citenamefont {Kayanuma},\ and\
  \citenamefont {H{\"a}nggi}}]{saito2007-dissipative}%
  \BibitemOpen
  \bibfield  {author} {\bibinfo {author} {\bibfnamefont {K.}~\bibnamefont
  {Saito}}, \bibinfo {author} {\bibfnamefont {M.}~\bibnamefont {Wubs}},
  \bibinfo {author} {\bibfnamefont {S.}~\bibnamefont {Kohler}}, \bibinfo
  {author} {\bibfnamefont {Y.}~\bibnamefont {Kayanuma}}, \ and\ \bibinfo
  {author} {\bibfnamefont {P.}~\bibnamefont {H{\"a}nggi}},\ }\href {\doibase
  10.1103/physrevb.75.214308} {\bibfield  {journal} {\bibinfo  {journal}
  {Physical Review B}\ }\textbf {\bibinfo {volume} {75}},\ \bibinfo {pages}
  {214308} (\bibinfo {year} {2007})}\BibitemShut {NoStop}%
\bibitem [{\citenamefont {Nalbach}\ and\ \citenamefont
  {Thorwart}(2009)}]{nalbach2009-landau}%
  \BibitemOpen
  \bibfield  {author} {\bibinfo {author} {\bibfnamefont {P.}~\bibnamefont
  {Nalbach}}\ and\ \bibinfo {author} {\bibfnamefont {M.}~\bibnamefont
  {Thorwart}},\ }\href {\doibase 10.1103/physrevlett.103.220401} {\bibfield
  {journal} {\bibinfo  {journal} {Physical Review Letters}\ }\textbf {\bibinfo
  {volume} {103}},\ \bibinfo {pages} {220401} (\bibinfo {year}
  {2009})}\BibitemShut {NoStop}%
\bibitem [{\citenamefont {Whitney}\ \emph {et~al.}(2011)\citenamefont
  {Whitney}, \citenamefont {Clusel},\ and\ \citenamefont
  {Ziman}}]{whitney2011-temperature}%
  \BibitemOpen
  \bibfield  {author} {\bibinfo {author} {\bibfnamefont {R.~S.}\ \bibnamefont
  {Whitney}}, \bibinfo {author} {\bibfnamefont {M.}~\bibnamefont {Clusel}}, \
  and\ \bibinfo {author} {\bibfnamefont {T.}~\bibnamefont {Ziman}},\ }\href
  {\doibase 10.1103/physrevlett.107.210402} {\bibfield  {journal} {\bibinfo
  {journal} {Physical Review Letters}\ }\textbf {\bibinfo {volume} {107}},\
  \bibinfo {pages} {210402} (\bibinfo {year} {2011})}\BibitemShut {NoStop}%
\bibitem [{\citenamefont {Haikka}\ and\ \citenamefont
  {M{\o}lmer}(2014)}]{haikka2014-dissipative}%
  \BibitemOpen
  \bibfield  {author} {\bibinfo {author} {\bibfnamefont {P.}~\bibnamefont
  {Haikka}}\ and\ \bibinfo {author} {\bibfnamefont {K.}~\bibnamefont
  {M{\o}lmer}},\ }\href {\doibase 10.1103/physreva.89.052114} {\bibfield
  {journal} {\bibinfo  {journal} {Physical Review A}\ }\textbf {\bibinfo
  {volume} {89}},\ \bibinfo {pages} {052114} (\bibinfo {year}
  {2014})}\BibitemShut {NoStop}%
\bibitem [{\citenamefont {Arceci}\ \emph {et~al.}(2017)\citenamefont {Arceci},
  \citenamefont {Barbarino}, \citenamefont {Fazio},\ and\ \citenamefont
  {Santoro}}]{arceci2017-dissipative}%
  \BibitemOpen
  \bibfield  {author} {\bibinfo {author} {\bibfnamefont {L.}~\bibnamefont
  {Arceci}}, \bibinfo {author} {\bibfnamefont {S.}~\bibnamefont {Barbarino}},
  \bibinfo {author} {\bibfnamefont {R.}~\bibnamefont {Fazio}}, \ and\ \bibinfo
  {author} {\bibfnamefont {G.~E.}\ \bibnamefont {Santoro}},\ }\href {\doibase
  10.1103/physrevb.96.054301} {\bibfield  {journal} {\bibinfo  {journal}
  {Physical Review B}\ }\textbf {\bibinfo {volume} {96}},\ \bibinfo {pages}
  {054301} (\bibinfo {year} {2017})}\BibitemShut {NoStop}%
\bibitem [{\citenamefont {Huang}\ and\ \citenamefont
  {Zhao}(2018)}]{huang2018-dynamics}%
  \BibitemOpen
  \bibfield  {author} {\bibinfo {author} {\bibfnamefont {Z.}~\bibnamefont
  {Huang}}\ and\ \bibinfo {author} {\bibfnamefont {Y.}~\bibnamefont {Zhao}},\
  }\href {\doibase 10.1103/physreva.97.013803} {\bibfield  {journal} {\bibinfo
  {journal} {Physical Review A}\ }\textbf {\bibinfo {volume} {97}},\ \bibinfo
  {pages} {013803} (\bibinfo {year} {2018})}\BibitemShut {NoStop}%
\bibitem [{\citenamefont {Makarov}\ and\ \citenamefont
  {Makri}(1994)}]{makarov1994-path}%
  \BibitemOpen
  \bibfield  {author} {\bibinfo {author} {\bibfnamefont {D.~E.}\ \bibnamefont
  {Makarov}}\ and\ \bibinfo {author} {\bibfnamefont {N.}~\bibnamefont
  {Makri}},\ }\href {\doibase 10.1016/0009-2614(94)00275-4} {\bibfield
  {journal} {\bibinfo  {journal} {Chemical Physics Letters}\ }\textbf {\bibinfo
  {volume} {221}},\ \bibinfo {pages} {482} (\bibinfo {year}
  {1994})}\BibitemShut {NoStop}%
\bibitem [{\citenamefont {Makri}(1995)}]{makri1995-numerical}%
  \BibitemOpen
  \bibfield  {author} {\bibinfo {author} {\bibfnamefont {N.}~\bibnamefont
  {Makri}},\ }\href {\doibase 10.1063/1.531046} {\bibfield  {journal} {\bibinfo
   {journal} {Journal of Mathematical Physics}\ }\textbf {\bibinfo {volume}
  {36}},\ \bibinfo {pages} {2430} (\bibinfo {year} {1995})}\BibitemShut
  {NoStop}%
\bibitem [{\citenamefont {Weiss}\ \emph {et~al.}(2008)\citenamefont {Weiss},
  \citenamefont {Eckel}, \citenamefont {Thorwart},\ and\ \citenamefont
  {Egger}}]{weiss2008-iterative}%
  \BibitemOpen
  \bibfield  {author} {\bibinfo {author} {\bibfnamefont {S.}~\bibnamefont
  {Weiss}}, \bibinfo {author} {\bibfnamefont {J.}~\bibnamefont {Eckel}},
  \bibinfo {author} {\bibfnamefont {M.}~\bibnamefont {Thorwart}}, \ and\
  \bibinfo {author} {\bibfnamefont {R.}~\bibnamefont {Egger}},\ }\href
  {\doibase 10.1103/physrevb.77.195316} {\bibfield  {journal} {\bibinfo
  {journal} {Physical Review B}\ }\textbf {\bibinfo {volume} {77}},\ \bibinfo
  {pages} {195316} (\bibinfo {year} {2008})}\BibitemShut {NoStop}%
\bibitem [{\citenamefont {Segal}\ \emph {et~al.}(2010)\citenamefont {Segal},
  \citenamefont {Millis},\ and\ \citenamefont
  {Reichman}}]{segal2010-numerically}%
  \BibitemOpen
  \bibfield  {author} {\bibinfo {author} {\bibfnamefont {D.}~\bibnamefont
  {Segal}}, \bibinfo {author} {\bibfnamefont {A.~J.}\ \bibnamefont {Millis}}, \
  and\ \bibinfo {author} {\bibfnamefont {D.~R.}\ \bibnamefont {Reichman}},\
  }\href {\doibase 10.1103/physrevb.82.205323} {\bibfield  {journal} {\bibinfo
  {journal} {Physical Review B}\ }\textbf {\bibinfo {volume} {82}},\ \bibinfo
  {pages} {205323} (\bibinfo {year} {2010})}\BibitemShut {NoStop}%
\bibitem [{\citenamefont {Nozi{\`e}res}\ and\ \citenamefont
  {Dominicis}(1969)}]{nozieres1969-singularities}%
  \BibitemOpen
  \bibfield  {author} {\bibinfo {author} {\bibfnamefont {P.}~\bibnamefont
  {Nozi{\`e}res}}\ and\ \bibinfo {author} {\bibfnamefont {C.~T.~D.}\
  \bibnamefont {Dominicis}},\ }\href {\doibase 10.1103/physrev.178.1097}
  {\bibfield  {journal} {\bibinfo  {journal} {Physical Review}\ }\textbf
  {\bibinfo {volume} {178}},\ \bibinfo {pages} {1097} (\bibinfo {year}
  {1969})}\BibitemShut {NoStop}%
\bibitem [{\citenamefont {Ng}(1995)}]{ng1995-x}%
  \BibitemOpen
  \bibfield  {author} {\bibinfo {author} {\bibfnamefont {T.-K.}\ \bibnamefont
  {Ng}},\ }\href {\doibase 10.1103/PhysRevB.51.2009} {\bibfield  {journal}
  {\bibinfo  {journal} {Physical Review B}\ }\textbf {\bibinfo {volume} {51}},\
  \bibinfo {pages} {2009} (\bibinfo {year} {1995})}\BibitemShut {NoStop}%
\bibitem [{\citenamefont {Ng}(1996)}]{ng1996-fermi}%
  \BibitemOpen
  \bibfield  {author} {\bibinfo {author} {\bibfnamefont {T.-K.}\ \bibnamefont
  {Ng}},\ }\href {\doibase 10.1103/physrevb.54.5814} {\bibfield  {journal}
  {\bibinfo  {journal} {Physical Review B}\ }\textbf {\bibinfo {volume} {54}},\
  \bibinfo {pages} {5814} (\bibinfo {year} {1996})}\BibitemShut {NoStop}%
\bibitem [{\citenamefont {Segal}\ \emph {et~al.}(2007)\citenamefont {Segal},
  \citenamefont {Reichman},\ and\ \citenamefont
  {Millis}}]{segal2007-nonequilibrium}%
  \BibitemOpen
  \bibfield  {author} {\bibinfo {author} {\bibfnamefont {D.}~\bibnamefont
  {Segal}}, \bibinfo {author} {\bibfnamefont {D.~R.}\ \bibnamefont {Reichman}},
  \ and\ \bibinfo {author} {\bibfnamefont {A.~J.}\ \bibnamefont {Millis}},\
  }\href {\doibase 10.1103/physrevb.76.195316} {\bibfield  {journal} {\bibinfo
  {journal} {Physical Review B}\ }\textbf {\bibinfo {volume} {76}},\ \bibinfo
  {pages} {195316} (\bibinfo {year} {2007})}\BibitemShut {NoStop}%
\bibitem [{\citenamefont {Chen}\ and\ \citenamefont
  {Xu}(2019)}]{chen2019-dissipative}%
  \BibitemOpen
  \bibfield  {author} {\bibinfo {author} {\bibfnamefont {R.}~\bibnamefont
  {Chen}}\ and\ \bibinfo {author} {\bibfnamefont {X.}~\bibnamefont {Xu}},\
  }\href {\doibase 10.1103/physrevb.100.115437} {\bibfield  {journal} {\bibinfo
   {journal} {Physical Review B}\ }\textbf {\bibinfo {volume} {100}},\ \bibinfo
  {pages} {115437} (\bibinfo {year} {2019})}\BibitemShut {NoStop}%
\bibitem [{\citenamefont {Makarov}\ and\ \citenamefont
  {Makri}(1995{\natexlab{a}})}]{makarov1995-control}%
  \BibitemOpen
  \bibfield  {author} {\bibinfo {author} {\bibfnamefont {D.~E.}\ \bibnamefont
  {Makarov}}\ and\ \bibinfo {author} {\bibfnamefont {N.}~\bibnamefont
  {Makri}},\ }\href {\doibase 10.1103/physreve.52.5863} {\bibfield  {journal}
  {\bibinfo  {journal} {Physical Review E}\ }\textbf {\bibinfo {volume} {52}},\
  \bibinfo {pages} {5863} (\bibinfo {year} {1995}{\natexlab{a}})}\BibitemShut
  {NoStop}%
\bibitem [{\citenamefont {Makarov}\ and\ \citenamefont
  {Makri}(1995{\natexlab{b}})}]{makarov1995-stochastic}%
  \BibitemOpen
  \bibfield  {author} {\bibinfo {author} {\bibfnamefont {D.~E.}\ \bibnamefont
  {Makarov}}\ and\ \bibinfo {author} {\bibfnamefont {N.}~\bibnamefont
  {Makri}},\ }\href {\doibase 10.1103/physrevb.52.r2257} {\bibfield  {journal}
  {\bibinfo  {journal} {Physical Review B}\ }\textbf {\bibinfo {volume} {52}},\
  \bibinfo {pages} {R2257} (\bibinfo {year} {1995}{\natexlab{b}})}\BibitemShut
  {NoStop}%
\bibitem [{\citenamefont {Makri}\ and\ \citenamefont
  {Wei}(1997)}]{makri1997-universal}%
  \BibitemOpen
  \bibfield  {author} {\bibinfo {author} {\bibfnamefont {N.}~\bibnamefont
  {Makri}}\ and\ \bibinfo {author} {\bibfnamefont {L.}~\bibnamefont {Wei}},\
  }\href {\doibase 10.1103/physreve.55.2475} {\bibfield  {journal} {\bibinfo
  {journal} {Physical Review E}\ }\textbf {\bibinfo {volume} {55}},\ \bibinfo
  {pages} {2475} (\bibinfo {year} {1997})}\BibitemShut {NoStop}%
\bibitem [{\citenamefont {Makri}(1997)}]{makri1997-stabilization}%
  \BibitemOpen
  \bibfield  {author} {\bibinfo {author} {\bibfnamefont {N.}~\bibnamefont
  {Makri}},\ }\href {\doibase 10.1063/1.473345} {\bibfield  {journal} {\bibinfo
   {journal} {The Journal of Chemical Physics}\ }\textbf {\bibinfo {volume}
  {106}},\ \bibinfo {pages} {2286} (\bibinfo {year} {1997})}\BibitemShut
  {NoStop}%
\bibitem [{\citenamefont {Segal}\ \emph {et~al.}(2011)\citenamefont {Segal},
  \citenamefont {Millis},\ and\ \citenamefont
  {Reichman}}]{segal2011-nonequilibrium}%
  \BibitemOpen
  \bibfield  {author} {\bibinfo {author} {\bibfnamefont {D.}~\bibnamefont
  {Segal}}, \bibinfo {author} {\bibfnamefont {A.~J.}\ \bibnamefont {Millis}}, \
  and\ \bibinfo {author} {\bibfnamefont {D.~R.}\ \bibnamefont {Reichman}},\
  }\href {\doibase 10.1039/c1cp20702d} {\bibfield  {journal} {\bibinfo
  {journal} {Physical Chemistry Chemical Physics}\ }\textbf {\bibinfo {volume}
  {13}},\ \bibinfo {pages} {14378} (\bibinfo {year} {2011})}\BibitemShut
  {NoStop}%
\bibitem [{\citenamefont {Simine}\ and\ \citenamefont
  {Segal}(2013)}]{simine2013-path}%
  \BibitemOpen
  \bibfield  {author} {\bibinfo {author} {\bibfnamefont {L.}~\bibnamefont
  {Simine}}\ and\ \bibinfo {author} {\bibfnamefont {D.}~\bibnamefont {Segal}},\
  }\href {\doibase 10.1063/1.4808108} {\bibfield  {journal} {\bibinfo
  {journal} {The Journal of Chemical Physics}\ }\textbf {\bibinfo {volume}
  {138}},\ \bibinfo {pages} {214111} (\bibinfo {year} {2013})}\BibitemShut
  {NoStop}%
\bibitem [{\citenamefont {Segal}(2013)}]{segal2013-qubit}%
  \BibitemOpen
  \bibfield  {author} {\bibinfo {author} {\bibfnamefont {D.}~\bibnamefont
  {Segal}},\ }\href {\doibase 10.1103/physrevb.87.195436} {\bibfield  {journal}
  {\bibinfo  {journal} {Physical Review B}\ }\textbf {\bibinfo {volume} {87}},\
  \bibinfo {pages} {195436} (\bibinfo {year} {2013})}\BibitemShut {NoStop}%
\bibitem [{\citenamefont {Agarwalla}\ and\ \citenamefont
  {Segal}(2017)}]{agarwalla2017-anderson}%
  \BibitemOpen
  \bibfield  {author} {\bibinfo {author} {\bibfnamefont {B.~K.}\ \bibnamefont
  {Agarwalla}}\ and\ \bibinfo {author} {\bibfnamefont {D.}~\bibnamefont
  {Segal}},\ }\href {\doibase 10.1063/1.4996562} {\bibfield  {journal}
  {\bibinfo  {journal} {The Journal of Chemical Physics}\ }\textbf {\bibinfo
  {volume} {147}},\ \bibinfo {pages} {054104} (\bibinfo {year}
  {2017})}\BibitemShut {NoStop}%
\bibitem [{\citenamefont {Makri}(1999)}]{makri1999-iterative}%
  \BibitemOpen
  \bibfield  {author} {\bibinfo {author} {\bibfnamefont {N.}~\bibnamefont
  {Makri}},\ }\href {\doibase 10.1063/1.479919} {\bibfield  {journal} {\bibinfo
   {journal} {The Journal of Chemical Physics}\ }\textbf {\bibinfo {volume}
  {111}},\ \bibinfo {pages} {6164} (\bibinfo {year} {1999})}\BibitemShut
  {NoStop}%
\end{thebibliography}
\end{document}